\documentclass[]{aa1}		 % for the 2-column author version

\usepackage{natbib}
\bibpunct{(}{)}{;}{a}{}{,}
\usepackage{lscape}
\usepackage{graphicx}
\usepackage{longtable}
\usepackage{supertabular}
\usepackage{txfonts}
\usepackage{color}

\definecolor{Red}{rgb}{1, 0, 0}
\definecolor{Green}{rgb}{0, 1, 0}
\definecolor{Blue}{rgb}{0, 0, 1}
\definecolor{Black}{rgb}{0, 0, 0}
\definecolor{White}{rgb}{1, 1, 1}
\definecolor{Grey}{rgb}{0.5, 0.5, 0.5}
\definecolor{Yellow}{rgb}{1, 1, 0}
\definecolor{Magenta}{rgb}{1, 0, 1}
\definecolor{Cyan}{rgb}{0, 1, 1}
\definecolor{Orange}{rgb}{0.8, 0.3, 0}
\definecolor{DarkGreen}{rgb}{0.2, 0.7, 0.1}
\definecolor{Pink}{rgb}{1, 0.4, 0.7}

\begin{document}

\title{JWST high-redshift galaxy constraints on warm and cold dark matter models}

\titlerunning{JWST constraints on WDM}

\authorrunning{U.~Maio \& M.~Viel}

\author{Umberto Maio,\inst{1}
		Matteo Viel,\inst{2,1,3,4}
	}

\offprints{U. Maio, e-mail: umberto.maio@inaf.it}

\institute{
    INAF-Italian National Institute of Astrophysics, Observatory of Trieste, via G.~Tiepolo 11, 34143 Trieste, Italy
	\and
    SISSA - International School for Advanced Studies, Via Bonomea 265, I-34136 Trieste, Italy
    	\and
	IFPU, Institute for Fundamental Physics of the Universe, Via Beirut 2, I-34151 Trieste, Italy
	\and
	INFN, Sezione di Trieste, Via Valerio 2, I-34127 Trieste, Italy
}

%\date{Received ...; ...}

%\abstract{}{}{}{}{}     %  5 {} tokens are mandatory:
\abstract
{Warm dark matter is a possible alternative to cold dark matter to explain cosmological structure formation.}
{We study the implications of the latest JWST data on the nature of dark matter.}
{We compare properties of high-redshift galaxies observed by JWST with hydrodynamical simulations, in the standard cold dark matter model and in warm dark matter models with a suppressed linear matter power spectrum}
{We find that current data are neither in tension with cold dark matter nor with warm dark matter models with $m_{\rm WDM} > 2$~keV, since they probe bright and rare objects whose physical properties are similar in the different scenarios. }
{We also show how two observables, the galaxy luminosity functions and the galaxy correlation function at small scales of faint objects, can be promising tools for discriminating between the different dark-matter scenarios.
Further hints may come from early stellar-mass statistics and galaxy CO emission.}

\maketitle

% =============================================================================

%***********************************************************************

\section{Introduction} \label{sect:introduction}

%**********************************************************************

\noindent
The early data release of the James Webb Space Telescope (JWST) has shown, for the first time, the existence of primordial galaxies in the very distant Universe, deep in the first half billion years.
These findings have profound implications on our understanding of primordial structure formation and can help pose tight constraints on the nature of dark matter.
While the common paradigm of structure formation relies on cold dark matter (CDM), that is matter that is non-relativistic at decoupling, alternative possibilities advocated to solve small-scale problems of the standard CDM scenario rely on the hypothesis that dark matter is made of warm particles that possess a small, non-negligible, thermal velocity, that is to say warm dark matter (WDM).
\\
This would affect the matter power spectrum producing a sharp decrease in power at relatively small scales  \citep{Colin2000, Bode2001} and would have significant implications on many astrophysical and cosmological observations (such as weak and strong lensing, clustering of galaxies, halo properties and their sub-haloes, intergalactic medium structures, and reionization).
In WDM scenarios the exact spectral damping depends on the particle physics model and if thermal WDM is considered this is parameterized by its mass, $m_{\rm WDM} $.
Available constraints point towards values varying from 1.5-2~keV (from Milky Way studies) to $\gtrsim 3-4$~keV (from Lyman-$\alpha$ forest flux investigations) ~\citep{Narayanan2000tp, Viel2005qj, Boyarsky2008xj, Maccio2010, Pacucci2013, Viel2013fqw, Palanque2020}.
However, if these limits are applied, then there is little room to disentangle WDM and CDM considering structure formation at low-redshift ($z < 4$)  \citep{Schneider2014}.
Thus, it becomes apparent that there is a unique window that needs to be explored to test dark matter (and even dark energy): the high-redshift  Universe~
\citep{MaioViel2015,Dayal2015,Corasaniti16,Corasaniti18,lapidanese, Rudakovskyi2021, lapi22,maio06, menci22, Kurmus22}.
\\
The first JWST observational programmes have detected ultra-high-redshift primordial galaxies, provided stellar masses and star formation estimates up to $z \simeq 12 $ 
\citep{Santini2022,Castellano2022,Naidu2022,Adams2022,Furtak2022, Rodighiero2022}, 
hinted at little dust content at those epochs and obtained direct metallicity measurements at $z \simeq 8$ \citep{Curti2022, CurtisLake2022}.
Furthermore, JWST data have invoked possible galaxy candidates at redshifts as high as $z \simeq 14 - 16$ \citep{Finkelstein2022, Donnan2022,Atek2022,Harikane2022, Bouwens2022}.
These are the first data probing the brightest galaxies at cosmic dawn and structure formation at such early cosmological epochs.
Therefore, they will allow us to pose, for the first time, constraints on the nature of dark matter based on the properties of the first galaxies in the infant Universe.
These data also raise obvious questions about early galaxy formation paths, challenge our ability to explain their existence at such primordial times \citep[see e.g.][]{behroozi}, and highlight the possibility that bright-galaxy number densities might be in excess in comparison to what is typically derived from fainter samples \citep{Bowler2020} or might be in tension with $\Lambda$CDM predictions 
\citep{boylan22,lovell22,naidu22,biagetti,haslbauer22}.
\\
Answering these questions is not a trivial task and requires detailed analysis of primordial structure formation, including the relevant chemical and physical mechanisms responsible for pristine-gas collapse and first-galaxy buildup.
To address these open issues, in the next sections, we perform several numerical simulations under different assumptions for CDM and WDM, and we contrast the recent JWST observational determinations against the predictions of our updated non-equilibrium numerical model applied to different dark-matter scenarios.
\\
Throughout this work, the used present-day density parameters 
for matter, cosmological constant, and baryons are
$\Omega_{\rm 0,m}=0.274$,   % Omega           0.274247  
$\Omega_{0,\Lambda}=0.726$, % OmegaLambda       0.725753
and $\Omega_{\rm 0,b}=0.0458$,
respectively, with a primordial helium mass fraction of 0.24.
The assumed expansion parameter normalized to 100~km/s/Mpc is $h=0.702$, while the adopted mass variance within an 8~Mpc radius and the spectral index are $\sigma_{8}=0.816$ and $n=0.968$ (consistent with WMAP data).
The present-day cosmological critical density is $ \rho_{0, crit}= 277.4 \, h^2 \, \rm M_\odot \, kpc^{-3}$.
\\
The paper is structured as follows. In Sect.~\ref{sect:simulations} we describe our methodology; in Sect.~\ref{sect:results} we discuss our main results about early-galaxy properties in CDM and WDM universes; and, finally, we summarize and conclude in Sect.~\ref{sect:conclusions}.
% 

%************************************************************************

\section{Simulations} \label{sect:simulations}

%************************************************************************

\noindent
We performed physics-rich and accurate simulations of cosmic structure formation in the high-$z$ Universe.
In more detail, we ran and analysed a number of numerical simulations based on the latest ColdSIM implementation for primordial galaxy formation by \cite{Maio2022}.
Briefly, that includes the whole variety of physical and chemical processes taking place in the primordial gas and leading to star formation in different cosmic environments 
\citep[see technical details in][]{Maio2022}.
It also represents a novel alternative modelling to what is commonly included in large numerical simulations
\citep[e.g.][]{kannan22}.
N-body and hydrodynamical calculations were performed by an extended version of the parallel code P-Gadget3  \citep{Springel2005}, modified to address cold-gas atomic and molecular physics over different cosmic epochs.
The implementation solved a self-consistent time-dependent non-equilibrium network of cosmic chemistry first-order differential equations (for the abundances of e$^-$, H, H$^+$, H$^-$, He, He$^+$, He$^{++}$, H$_2$, H$^+_2$, D, D$^+$, HD, and HeH$^+$ species), accounting for ionization, recombination, and dissociation processes \citep{Maio2007, Maio2010, Maio2011}.
It additionally included star formation, stellar evolution, and heavy-element enrichment of He, C, N, O, Ne, Mg, Si, S, Ca, Fe, among others from SN~II, AGB and SN~Ia phases \citep{Tornatore2007}.
Chemical properties were evaluated by considering any stellar particle as a single stellar population with a Salpeter initial mass function and mass-dependent stellar lifetimes. Metals were spread over the neighbour particles and mixing was mimicked through SPH kernel smoothing. Mechanical and chemical feedback were incorporated as well and they are responsible for regulating star formation over cosmological epochs.
Primordial gas cools via pristine H$_2$ formation (H$^-$, H$_2^+$ and three-body) channels, while enriched media host metallicity-dependent H$_2$ dust grain catalysis and cooling  \citep{Maio2022}.
The redshift-dependent dust grain temperature was estimated from the energy balance between CMB radiation and dust grain power-law emission with slope $\beta = 2$. The effects of the establishment of a UV background resulting from the ongoing structure evolution were taken into account with corresponding HI and H$_2$ shielding corrections and complemented with photo-electric and cosmic-ray heating in star-forming sites \citep[see][for theoretical details and parameter studies]{Maio2022}.
In this way, the model can correctly reproduce the amounts of atomic and molecular gas at different epochs and explain corresponding depletion times as inferred from observational data at $z < 7$ \citep{Maio2022}.
\\
To assess the effects of baryon physics and possible degeneracies, we considered a fiducial implementation fully including all the above-mentioned physical processes, as well as a basic implementation including only the two fundamental (H$^-$ and H$^+_2$) channels for H$_2$ evolution and cooling at primordial epochs.
We used a reference simulation box size of $10\,\rm\,Mpc / {\it h}$ with $2\times 512^{3}$ particles, giving gas and dark-matter particle masses of
$M_{\rm gas} \simeq  7.9 \times 10^4\, \rm M_{\odot}/ {\it h}$ and 
$M_{\rm dm} \simeq 4.7 \times 10^5 \, \rm M_{\odot} / {\it h}$.
Our fiducial implementation was employed to run higher-resolution (HR) simulations of initial $2\times 1000^{3}$ gas and dark-matter particles with $M_{\rm gas} \simeq 10^4\, \rm M_{\odot}/ {\it h}$ and $M_{\rm dm} \simeq 6 \times 10^4 \, \rm M_{\odot} / {\it h}$, as well as a CDM large-box (LB) run of initial $2 \times 1000^3$ gas and dark matter particles in a 50~Mpc/$h$-side volume.
The simulations were started at $z = 99$, with initial conditions generated on a regular grid using the CAMB transfer function \citep{Lewis1999bs} and CDM and WDM power spectra with $m_{\rm WDM} = 2 $ and $3\,\rm keV$.
Thanks to our resolution we were able to properly resolve gravitational collapse and molecular catastrophic cooling down to Jeans masses of the order of $ \sim 10^6-10^8 \,\rm M_\odot $. These values are typical for the regimes of interest here.
\\
We note that several implementations of cosmic structure formation are available in the literature; however, most of them are based on local-Universe calibrations and do not consider any proper modelling of primeval epochs.
Instead, for this work, we performed ab initio (down to molecular interactions)
CDM and WDM simulations of the pristine-gas collapse and consequent birth of the first stars and galaxies by including all the detailed physical and chemical processes involved at such times.
This makes our findings particularly complete and reliable.
As a follow-up to our previous work \citep{MaioViel2015}, which nevertheless lacks the $ m_{\rm WDM } = 2\,\rm keV $ case, for this work we have been able to run simulations at much higher resolution, with a more advanced and accurate numerical implementation and by exploring a wider parameter space.
In the next, we show our main results and their implications on the nature of dark matter in light of the first JWST data release.

%*****************************************************************************

\section{Results} \label{sect:results}

%*****************************************************************************

\begin{figure*}
  \centering
  \includegraphics[width=0.32\textwidth]{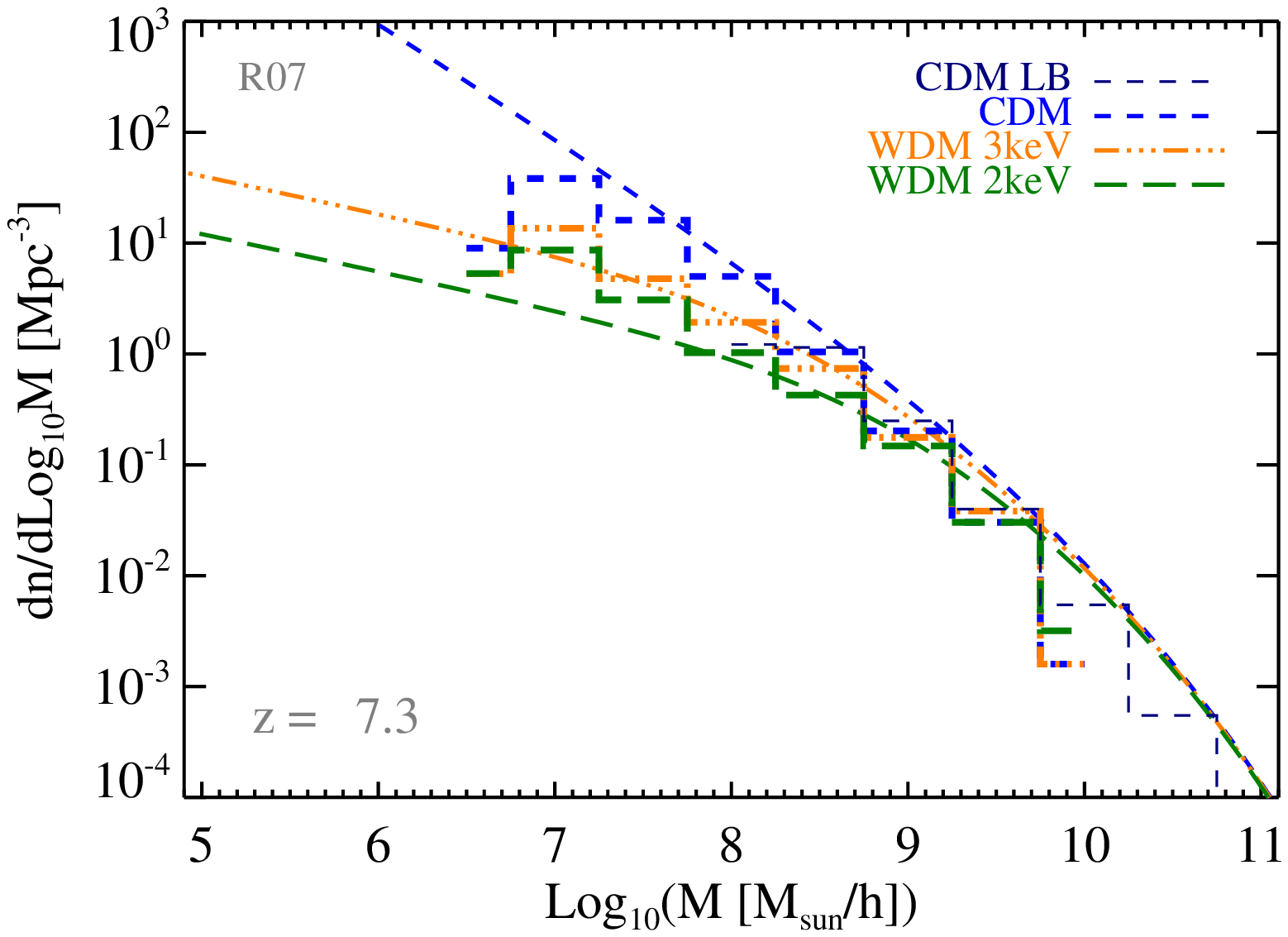}  % fix dot-trouble by ".ps"
  \includegraphics[width=0.32\textwidth]{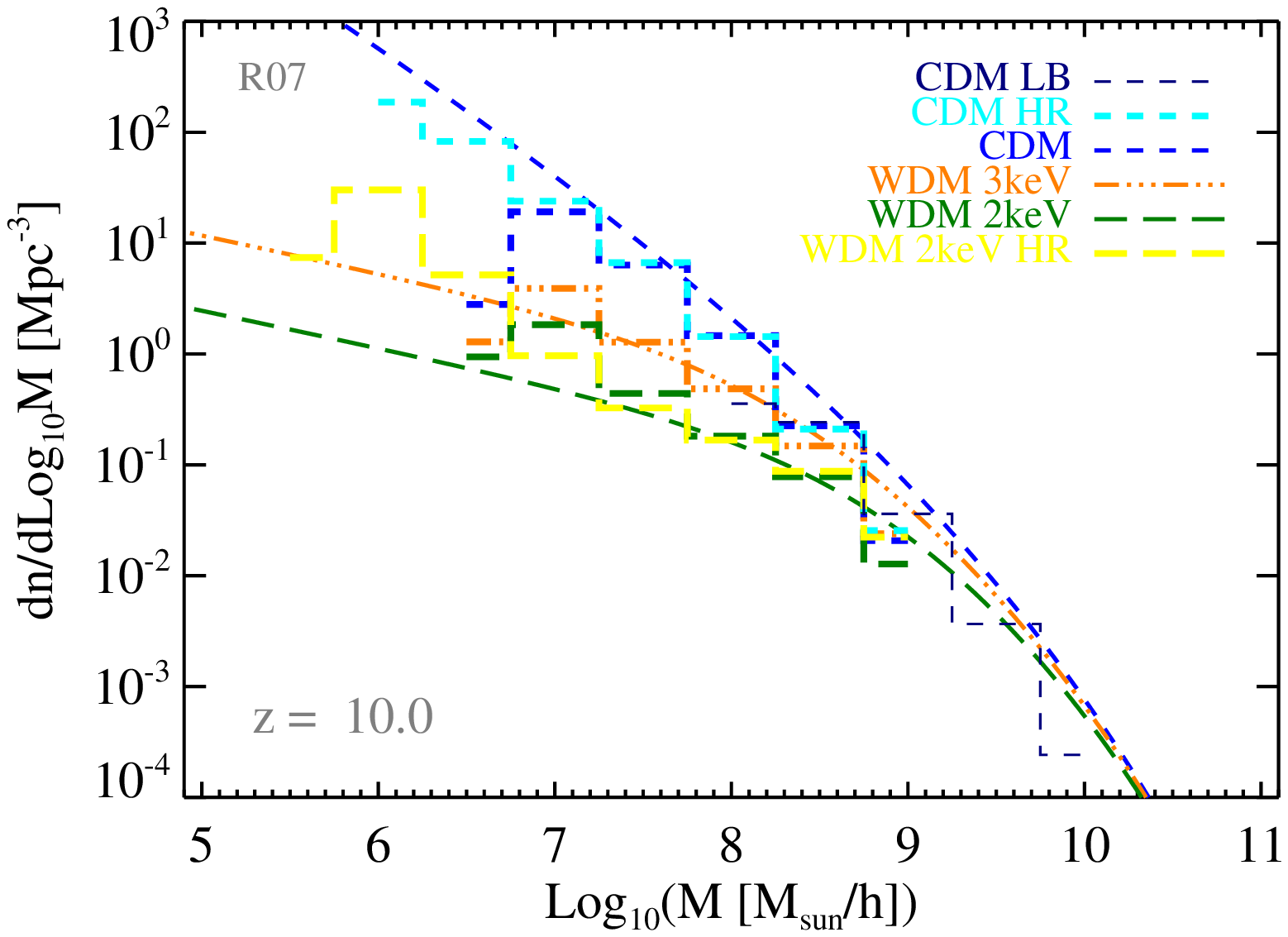}
  \includegraphics[width=0.32\textwidth]{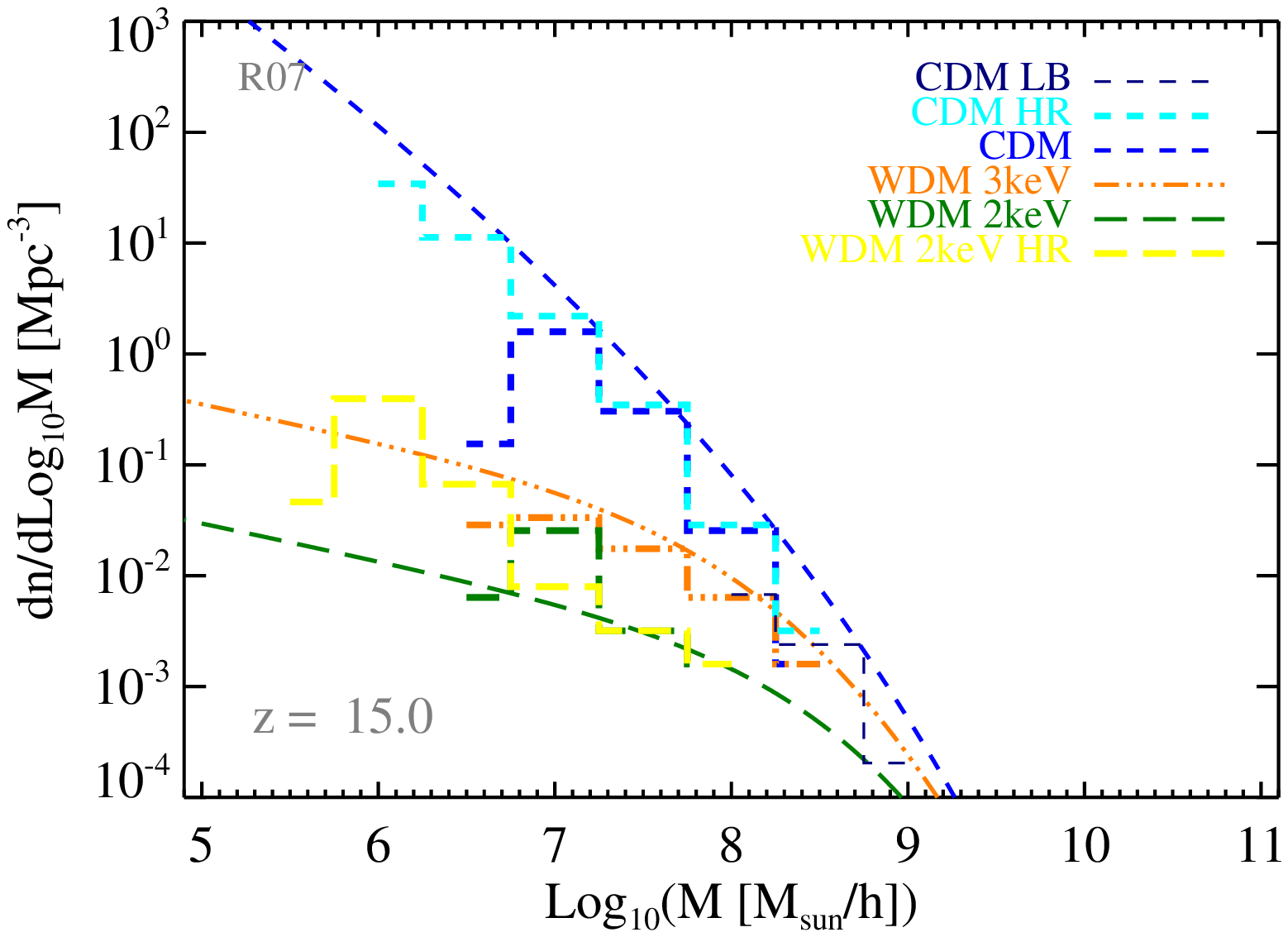}
\caption{Halo mass functions for the different dark-matter models at different redshifts. Numerical simulation fiducial results (histograms) are overplotted on the corresponding analytical expectations (lines) by \cite{R2007}. These mass functions have been derived by considering hydro-dynamical simulations and are thereby incorporating baryonic effects.
}
  \label{fig:mfR}
\end{figure*}

\begin{figure}
  \centering
  \includegraphics[width=0.45\textwidth]{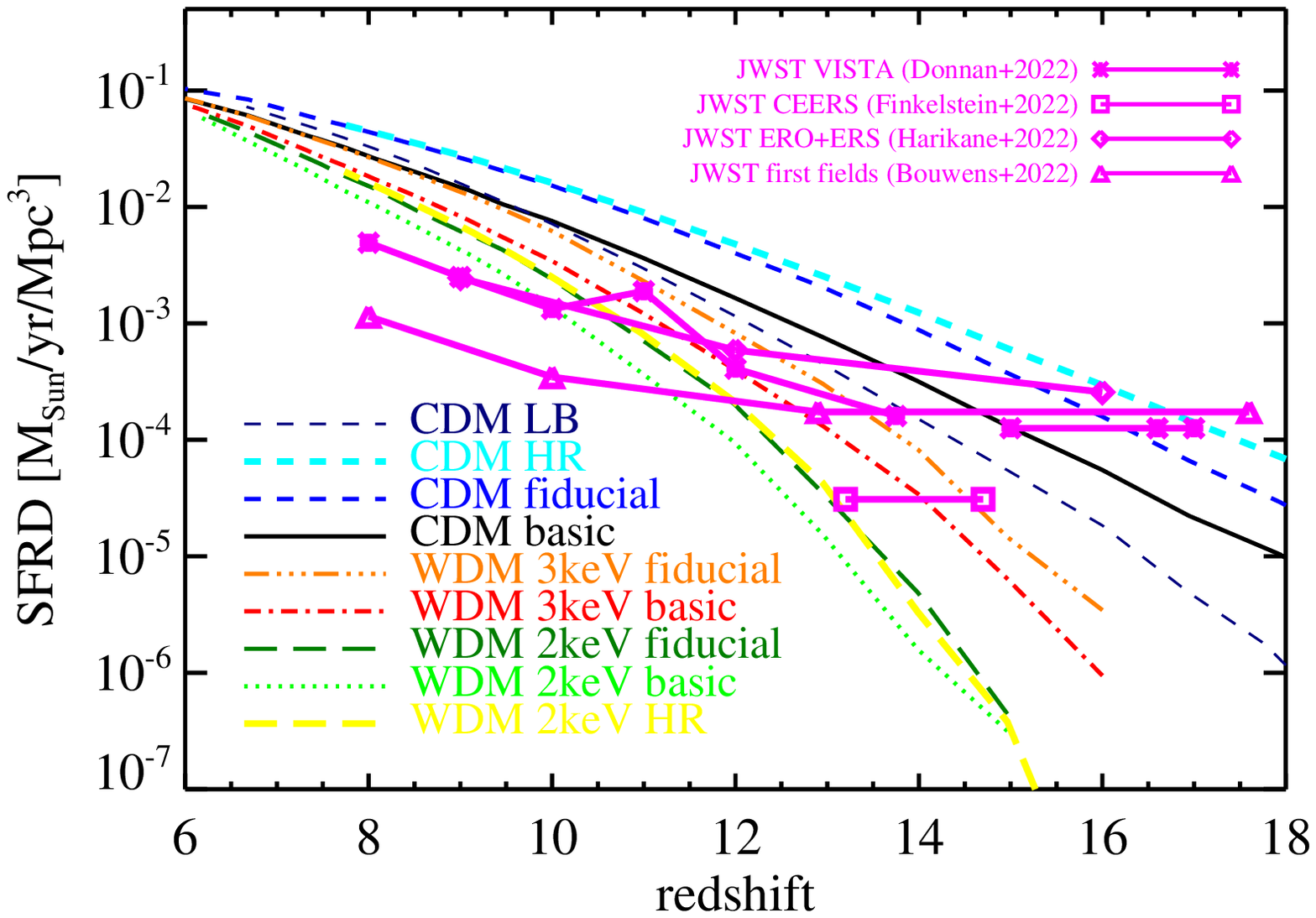} 
\caption{Redshift evolution of the cosmological SFR density derived from different models and compared to JWST determinations at $6 \lesssim z \lesssim 17 $ \citep{Donnan2022,Finkelstein2022,Harikane2022,Bouwens2022}. 
}
  \label{fig:SFRD}
\end{figure}

\noindent
As a sanity check of the proper-behaviour of our simulation results, the simulated mass functions for the considered dark-matter models of the fiducial runs are plotted at different redshifts and compared to corresponding analytical expectations \citep{R2007} in fig.~\ref{fig:mfR}.
The trends of the different models converge towards the high-mass end, while there are clear departures at lower masses. This is a direct consequence of the suppression of low-mass haloes determined by the WDM transfer functions.
We notice that discrepancies among the three cases become milder at lower $ z $, while they are larger at $ z \gtrsim 10 $, when matter non-linearities have not had time to grow significantly.
For this reason the high-$z$ regime is an exceptional probe of the nature of dark matter.
Results from the CDM LB and HR runs are in excellent agreement with analytical predictions and with the reference CDM run.
\\
In fig.~\ref{fig:SFRD}, the cosmic star formation rate density (SFRD) is shown for both simulation results and first JWST Early Release Observations/Science programmes (ERO/ERS), including CEERS, GLASS, and SMACS~0723.
It is striking that the fiducial CDM trend is consistent with all the constraints available in the $ 7 \lesssim z \lesssim 16 $ range. Changes in the details of the implementation (fiducial versus basic) or increasing resolution, as for CDM HR, would not affect this result (the CDM LB run has a coarser resolution and clearly misses small objects; nevertheless, it hosts larger structures and is crucial for estimating number densities of the rarest objects).
The case of WDM with $ m_{\rm WDM } = $~3~keV is similar to the CDM, although the high-$z$ behaviour would be in tension with the claimed detection (to be confirmed) at $ z \simeq 16$.
We must note in this respect that high-$z$  measurements (such as star formation rates and luminosity distribution slopes) are degenerate with dust mass estimates in the selected objects. Hence, candidates at extremely high redshift (such as the $ z \simeq 16$ one) could be explained by dusty lower-$z$ counterparts and this would lead to ambiguous conclusions
\citep{Stefanon2021,Finkelstein2022,Furtak2022}.\\
The picture for 2~keV WDM is much clearer, with predictions being below all the observational lower limits at $ z \gtrsim 10 $ both in the reference fiducial and basic runs as well as in the HR run.
This is an insightful finding because, thanks to the newly acquired JWST observational data and by matching them with our state-of-the-art theoretical predictions, we could be able to put the 2~keV WDM model under stress
\citep[the 2~keV WDM mass is not consistent with Lyman-$\alpha$ forest observations, although has been claimed to be a possible lower bound for HST UV luminosity functions;][]{irsic17, Rudakovskyi2021}.
A very important point we want to make is that a proper modelling of primordial structure formation (such as the one we perform here) is fundamental to obtain meaningful results.
In this respect, even resolution issues seem to have a lesser impact than the details of the physical and chemical modelling.
Given the trends above, it is tempting to conclude that 2-3~keV WDM is ruled out. However, it must be noticed that global quantities extracted from observations of a few objects require extrapolation of physical quantities (such as the mass fraction in stars or luminosities) in regimes which are not probed by the data. For this reason in the rest of this work we turn our focus to the physical quantities associated to galaxies.
\\

\begin{figure}
  \centering
  \includegraphics[width=0.45\textwidth]{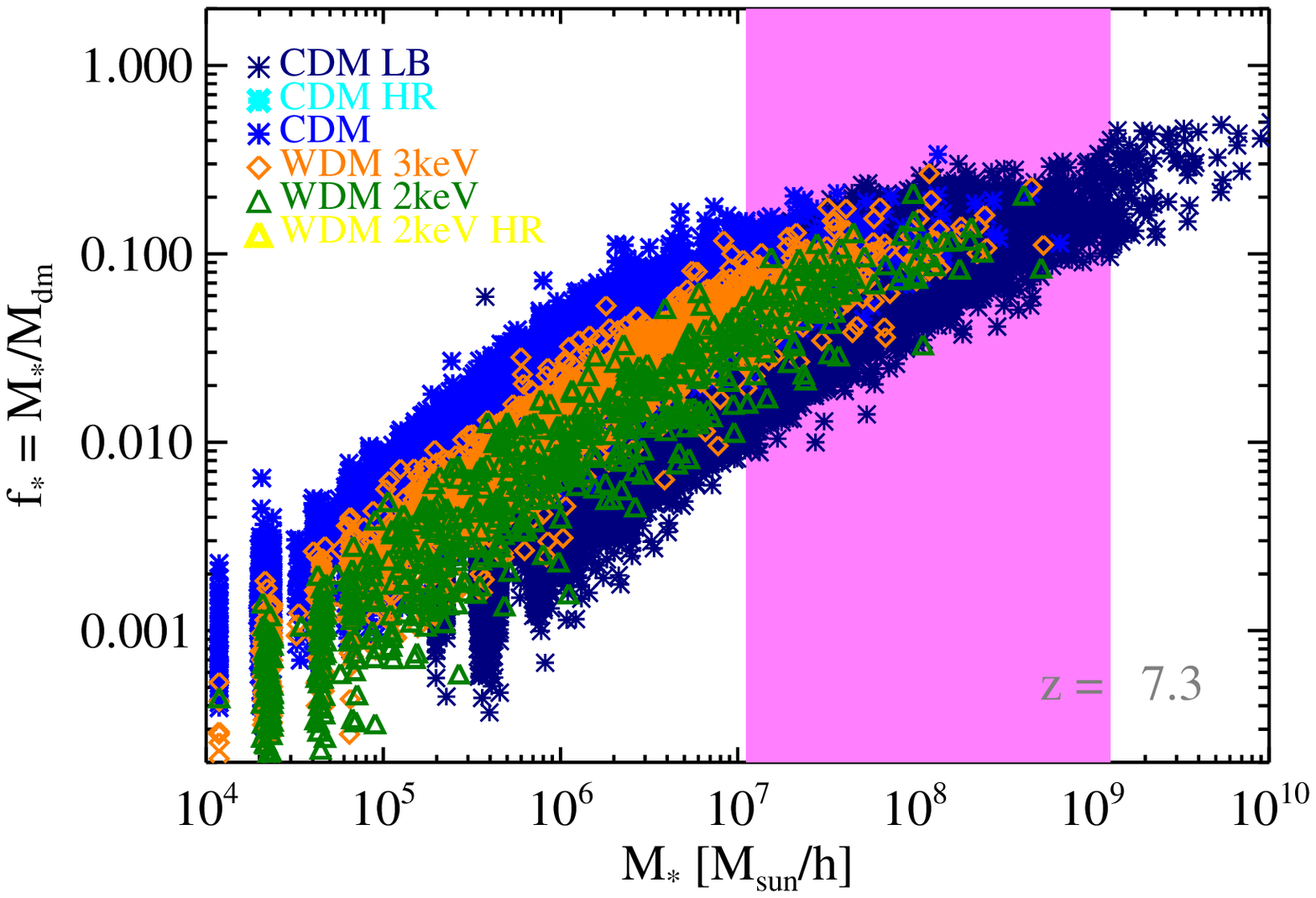}\\
  \vspace{-0.5cm}
  \includegraphics[width=0.45\textwidth]{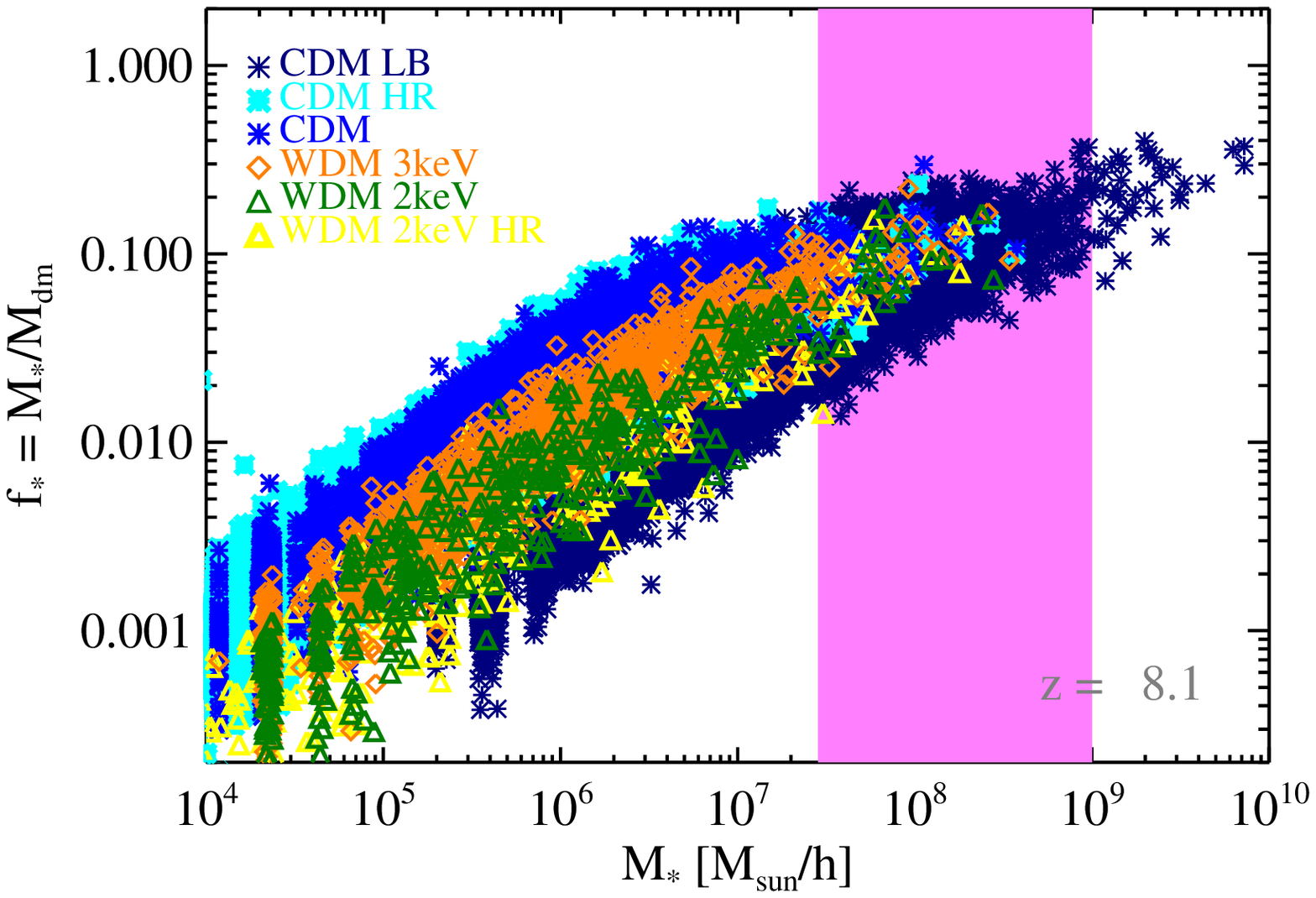}\\
  \vspace{-0.5cm}
  \includegraphics[width=0.45\textwidth]{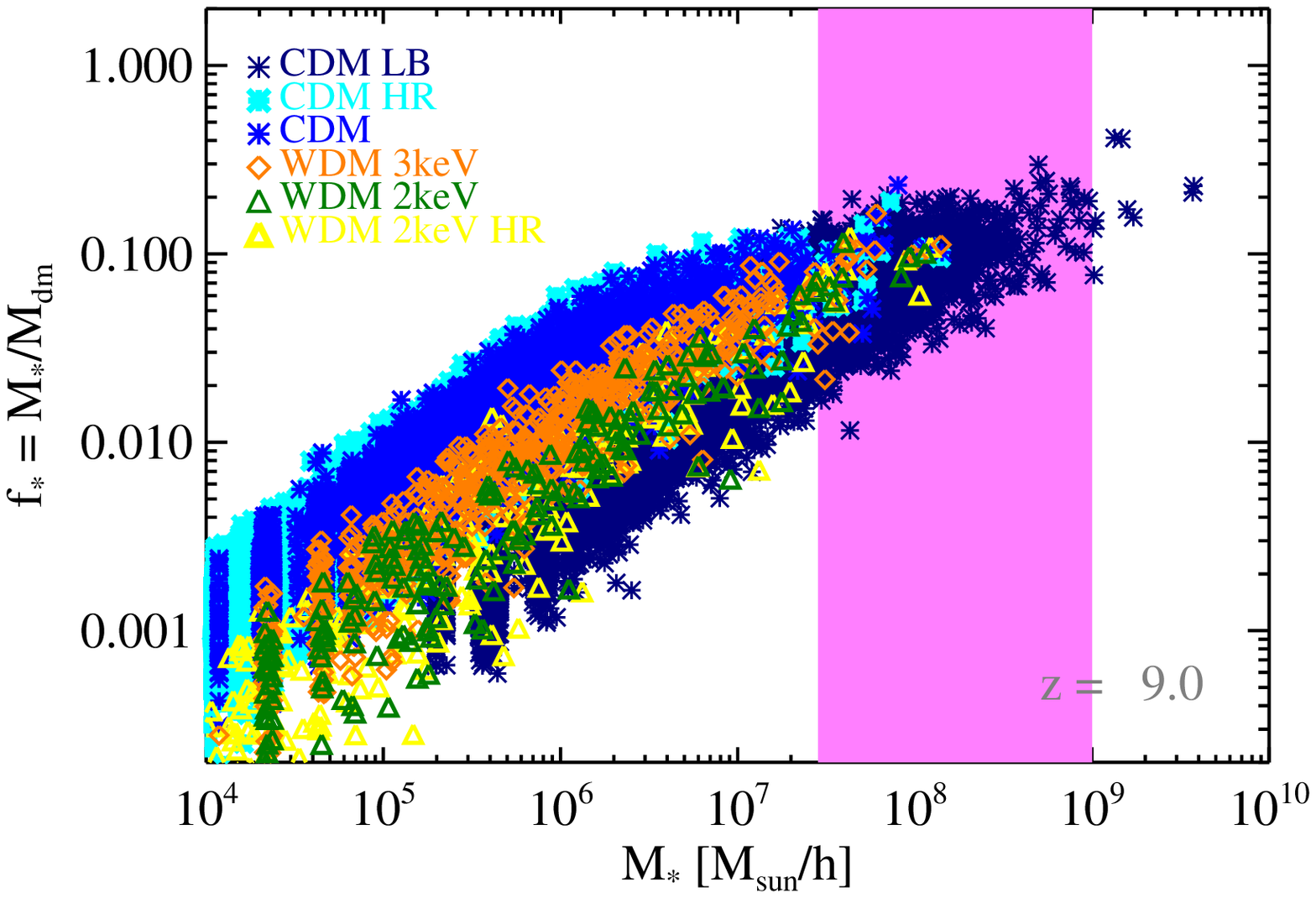}\\
  \vspace{-0.5cm}
  \includegraphics[width=0.45\textwidth]{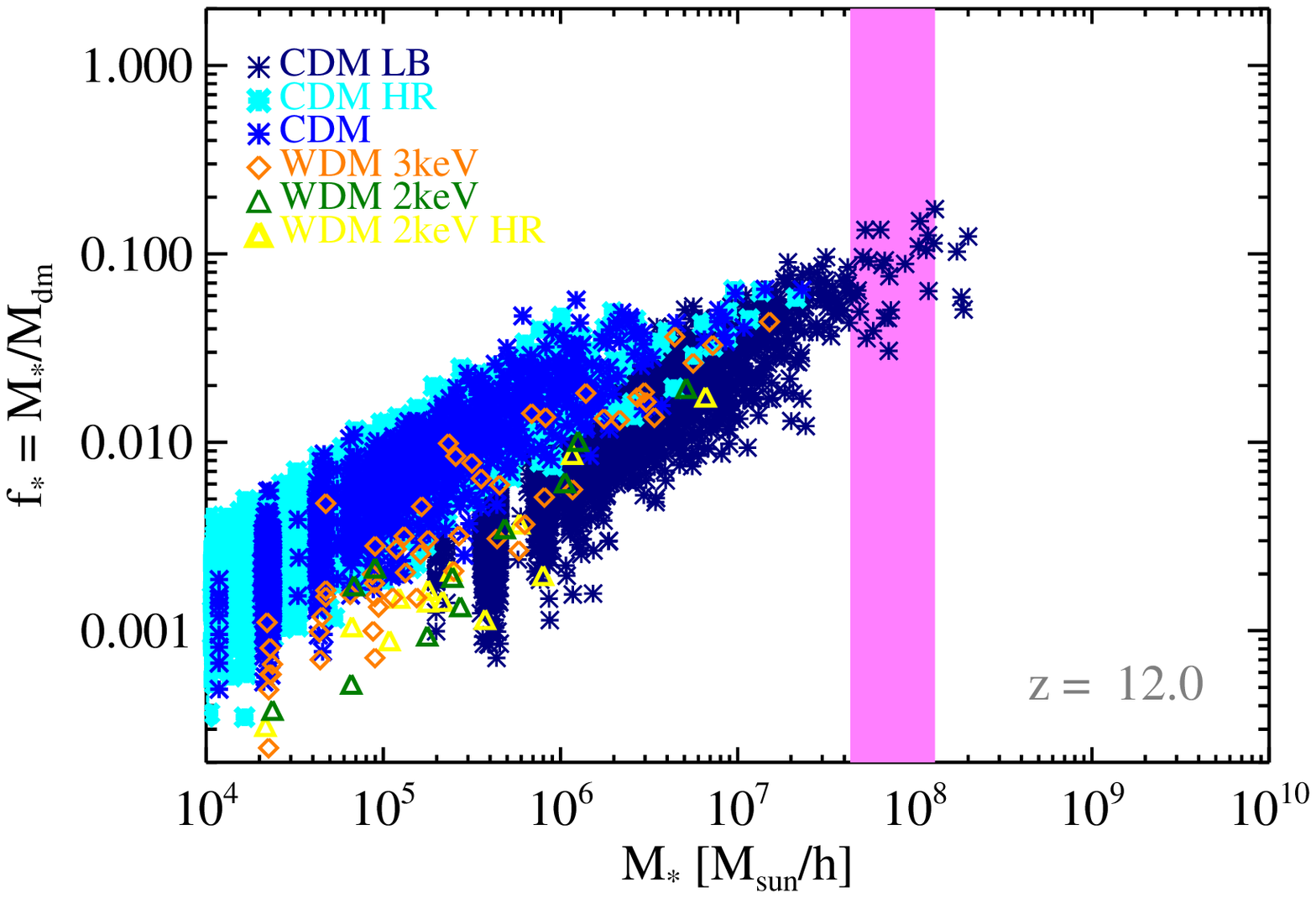}
\caption{Stellar mass fraction as a function of stellar mass for different models and resolutions at different redshifts, as indicated by the legend. The shaded area corresponds to the range of stellar masses inferred by GLASS-JWST ERS between $ z \simeq $ 6.9 and 12.1  \citep{Santini2022}.
}
  \label{fig:fstarMstar}
\end{figure}

\noindent
Basic early-halo properties are reminiscent of dark-matter features, as demonstrated by the mass distribution of the stellar fraction, $ f_\star $, plotted in fig.~\ref{fig:fstarMstar}, where numerical predictions are confronted with early JWST stellar-mass determinations \citep{Santini2022} at the same cosmological epoch.
Considering the reference runs, the CDM case shows a variety of masses and $ f_\star $ values that increase almost linearly already by the first half billion years, while WDM models predict a scarcer number of primordial structures (most notably in the 2~keV scenario) with smaller $ f_\star$ values.
These objects are just forming and have young lifetimes.
The mass-$ f_\star $ trends in WDM scenarios are caused by the 
cosmic gas accumulated over time in WDM  haloes (unable to callapse at early epochs due to matter power suppression) that, as soon as their masses/dimensions reach the spectral cutoff, suddenly become unstable and form stars.
Discrepancies are clearly visible in the 2~keV case, while the 3~keV displays an intermediate distribution between the latter and the CDM.
In general, smaller haloes are more sensitive to the underlying nature of dark matter; instead, the high-mass behaviour tends to converge both in mass and in $ f_\star $.
Observed stellar-mass estimates (shaded areas in the panels) rely on available calibrations and could bear uncertainties up to a factor of $\sim 3$ \citep{Stefanon2021}; although, in the long run, JWST is expected to improve that \citep[][]{Glass-Jwst2022}.
Our results are in agreement with latest determinations and can be considered solid and sound.
We are able to reach such a conclusion because of our accurate physical modelling explicitly focussed on high-$z$ galaxy formation.
For the sake of completeness, we also show the trends for the HR and LB runs.
We see that HR trends are in line with the lower-resolution ones, and thus our results are converged and do not suffer from numerical artefacts.
Also the CDM LB simulation, which is able to probe larger masses, displays comparable values for the stellar mass fraction,  despite its coarser resolution.
\\
\begin{figure}
  \centering
  \includegraphics[width=0.45\textwidth]{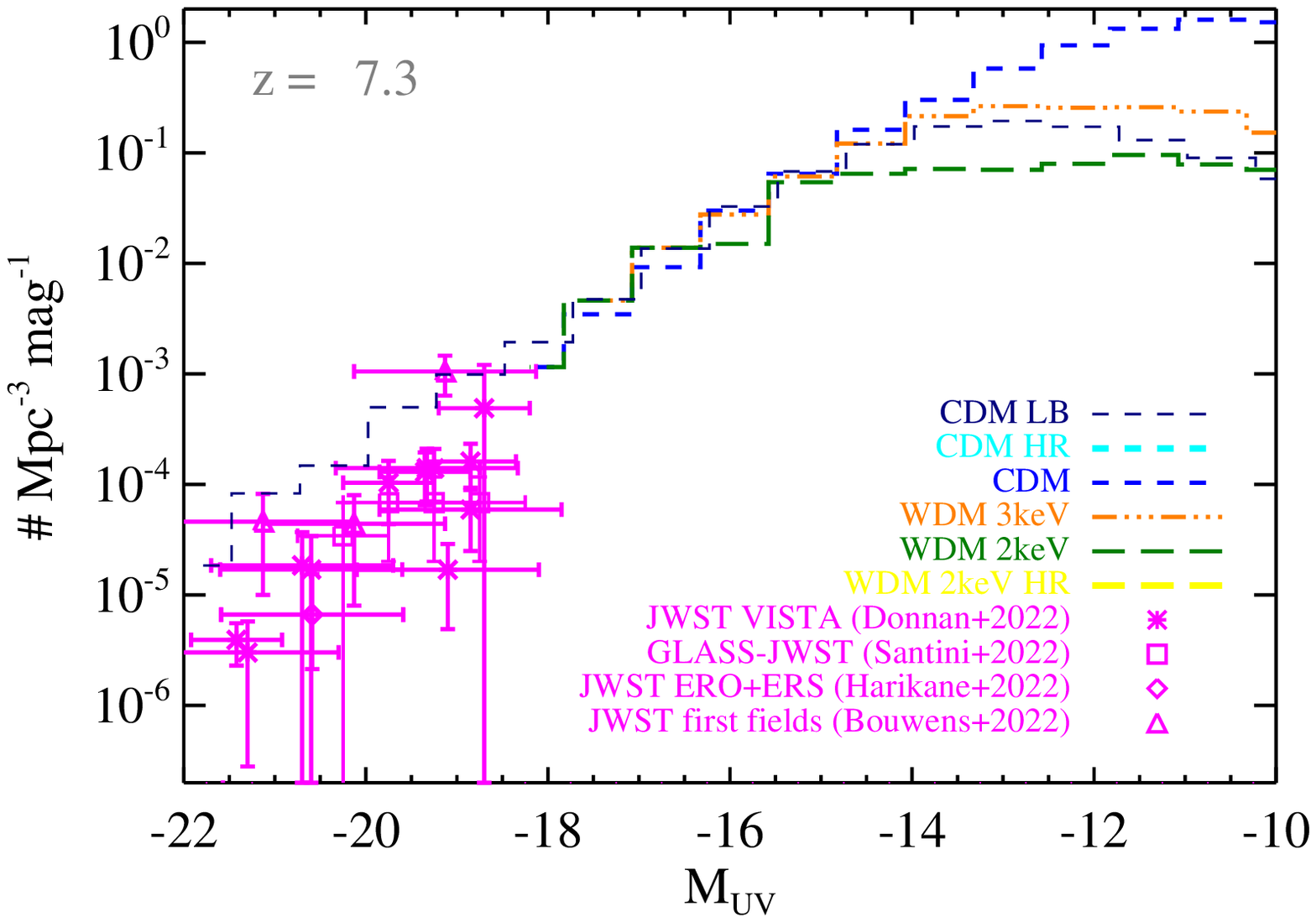}\\
  \vspace{-0.5cm}
  \includegraphics[width=0.45\textwidth]{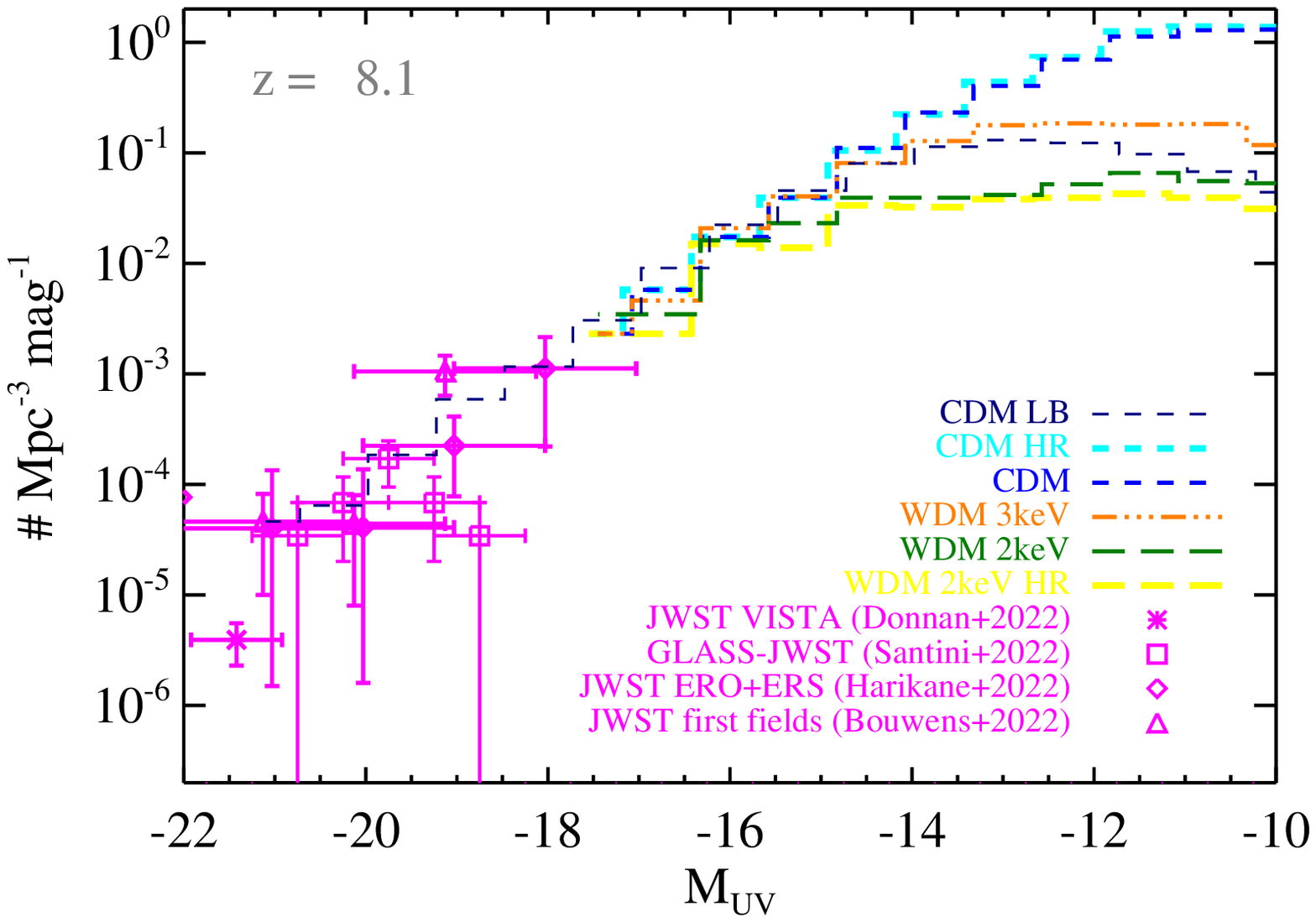}\\
  \vspace{-0.5cm}
  \includegraphics[width=0.45\textwidth]{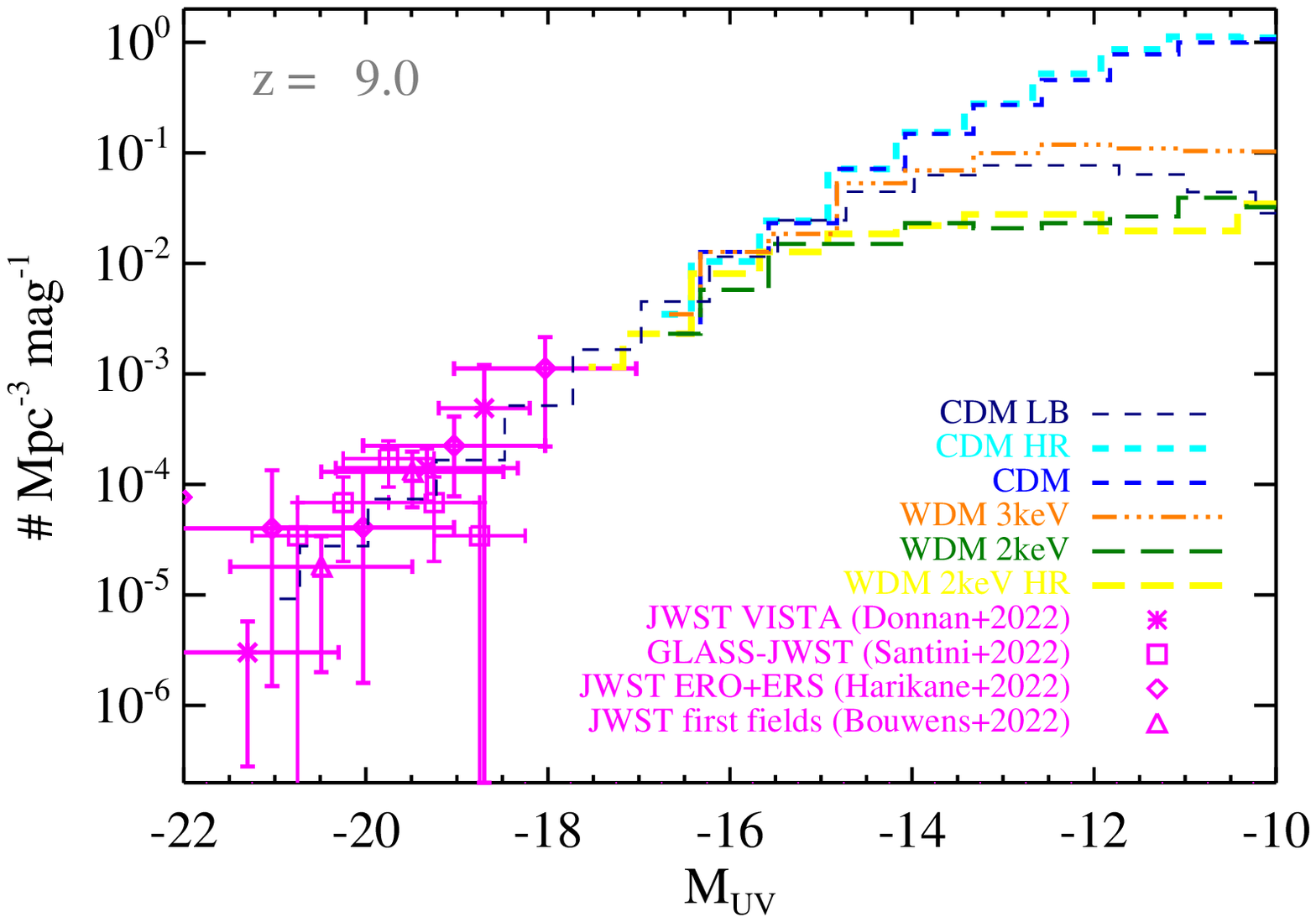}\\
  \vspace{-0.5cm}
  \includegraphics[width=0.45\textwidth]{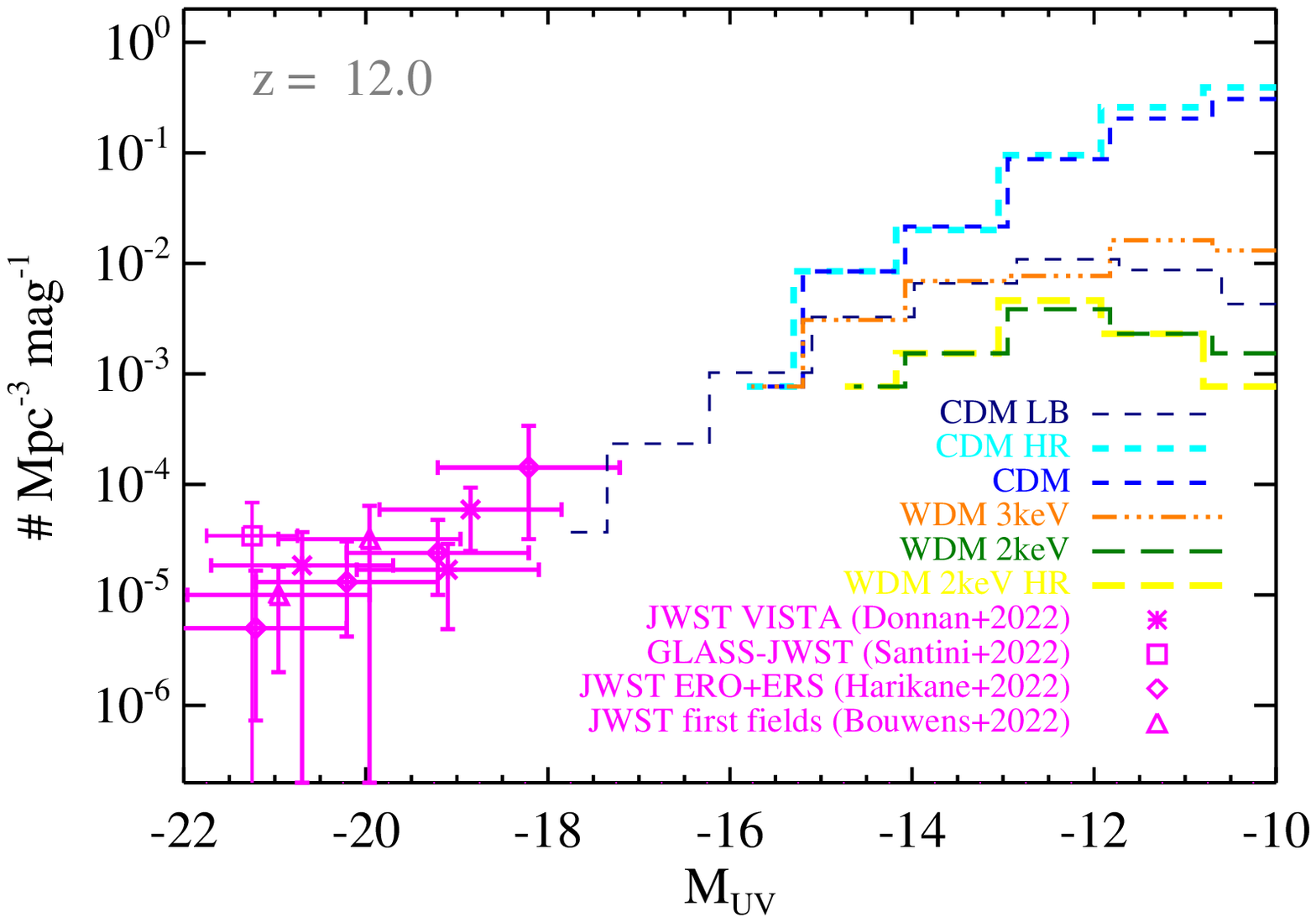}
\caption{Number of galaxies per unit co-moving volume and magnitude as a function of the UV magnitude at different $z$, for different dark-matter models and simulation resolutions. Theoretical results are compared to the latest JWST observational determinations \citep{Donnan2022, Santini2022, Harikane2022, Bouwens2022}.
}
  \label{fig:magUV}
\end{figure}

\noindent
In fig.~\ref{fig:magUV}, the UV luminosity functions at different redshifts and for the CDM and WDM dark-matter models are shown. The UV luminosity function was computed as the 
% number of galaxies per unit co-moving volume and magnitude.
number of galaxy per unit co-moving volume and magnitude.
% --> language editor: "galaxy per unit co-moving... since modifiers should not be pluralized."
UV magnitudes, M$_{\rm UV}$, are given in the AB system and were computed from simulation outputs by following standard prescriptions for simple stellar populations \citep{BC2003}.
The lack of simulated data below $ \sim 10^{-3} \,\rm Mpc^{-3} \, mag^{-1}$ (where the CDM and WDM trends have roughly converged) for the reference runs is due to the limited volume size.
This issue is alleviated by the LB results, which reach values of the order of $ 10^{-5} \,\rm Mpc^{-3}  mag^{-1} $ thanks to the larger volume sampled.
The theoretical trends for different resolutions are in excellent agreement at all $\rm M_{UV}$ values and have been compared to the latest JWST observational determinations \citep{Santini2022, Donnan2022, Harikane2022, Bouwens2022}.
As expected from the previous considerations, the faint end is severely affected by the differences in the input matter power spectra at small scales \citep[e.g.][]{barkana01, MaioViel2015, Rudakovskyi2021, lapi22}, while the bright end (at $M_{\rm UV} \lesssim -15 $) is substantially unchanged (this also applies to determinations at $z \simeq 16$, for which the trend is similar to the $z\simeq 12 $ panel, albeit with poorer statistics) and features a converging behaviour of the different models.
In this latter case, individual high-$z$ detections are accompanied by large errors and do not have any distinctive impact for CDM versus WDM nature.
At very early times ($z \simeq 9$-12) WDM and CDM models are clearly distinguishable, with differences in the number density of visible structures spanning almost 3 orders of magnitude at the faint end. Over cosmological evolution, the WDM models tend to catch up with the CDM and discrepancies in the faint-end distribution get reduced to only 1 dex at $ z\simeq 7.3$-8.
Thus, detection of luminous objects at early times is not sufficient to pose stringent constraints on the nature of dark matter via luminosity distributions and dim high-$z$ sources need to be investigated.
By comparing simulation results to observational data, we can conclude that galaxy abundances at the bright end of the luminosity function, where different models have already converged, is in line with predictions and can be explained through the fundamental physical processes taking place during early galaxy buildup, that is: rapid H$_2$-driven cosmic-gas collapse and stellar emission from young (a few tens of millions of years) stellar populations \citep{Mason2022,Inayoshi2022,Ferrara2022}.
\\
This means that, at least for $ z <12 $, there is clearly no excess of bright structures in the primordial Universe with respect to any model considered here
\citep[contrary to what is suggested via semi-analytic arguments by e.g.][]{Bowler2020,boylan22,lovell22,naidu22}), 
as is clearly visible from the behaviour in the figure.
At higher $z$ we are limited by the large observational errors ($M_{\rm UV }$ between $ -20$ and $-18$ for luminosity function values of $ 10^{-5} - 10^{-4} \,\rm Mpc^{-3} \, mag^{-1}$ and a few dex errors in the luminosity function at $ M_{\rm UV } < -20$), but the inferred trend seems to be consistent with the observational range shown in the bottom panel.
\\

\begin{figure}
  \includegraphics[width=0.45\textwidth]{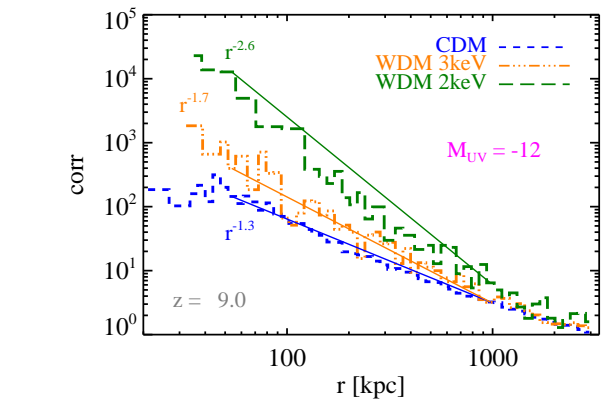}
\caption{Correlation functions of galaxies with UV magnitudes  $ M_{\rm UV} \simeq -12$ at $z = 9$.
Straight lines represent interpolated power-law trends with slopes of 
$ -1.31$, $ -1.66$, and $-2.62$
% slope cdm:       -1.3113080
% slope wdm3:       -1.6635471
% slope wdm2:       -2.6207266
for CDM, 3 keV~WDM, and 2 keV~WDM fiducial results, respectively.
}
  \label{fig:corrUV}
\end{figure}

\noindent
In fig.~\ref{fig:corrUV}, the two-point correlation functions of galaxies with UV magnitude around $ M_{\rm UV }=-12 $ (data selected between -13 and -10.5) at $ z=9 $ for the reference fiducial runs in the different dark-matter models is shown.
These objects are small, weakly star-forming structures that feature a bursty nature (i.e. large SFR per unit stellar mass), due to their young ages.
For a sample of $ N $ objects in a box with volume $ V $, the two-point correlation function, $\xi(r)$, is defined as the excess probability with respect to a uniform distribution of finding two objects at a given distance $r$.
It is computed by evaluating the number density of pairs in any given spherical shell with volume $ V_{\rm shell}(r) $, 
$ n_{\rm pairs}(r) = N_{\rm pairs}(r) / V_{\rm shell}(r) $, divided by $N$ and normalized by the expectation value of the corresponding uniform distribution with $ n = N/V$, i.e.:
$ 1 + \xi(r) 
= n_{\rm pairs}(r) / N / n  
= N_{\rm pairs }(r) V / \left[N^2 V_{\rm shell}(r) \right] $.\footnote{
This definition is adopted here and guarantees that results converge to unity (i.e. no excess probability, $\xi \simeq 0$) at large distances and is not affected by the size $N$ of the sample considered.}
\\
In fig.~\ref{fig:corrUV} at small radii (below a few $ \sim 10^2$~kpc), there are clear differences among the different models.
The CDM model presents a shallow behaviour (slope of -1.3) because of the more broadly distributed structures. The 2~keV WDM model, despite predicting fewer galaxies, features a steeper trend (slope of -2.6) as a result of the suppression of low-mass objects and galaxy satellites at large $r$. Therefore, WDM structures are much more clustered (up to a factor of 100) below the Mpc scale, as compared to CDM simulations. The 3~keV WDM has an intermediate behaviour (with a slope of about -1.7).
The  small-scale correlation function is thus sensitive to the different bias of the galaxy population, with WDM galaxies being more biased with regard to the corresponding CDM case.
This picture is qualitatively similar at different redshifts and $ M_{\rm UV}$, although at later times and for brighter magnitudes the models tend to converge. 
\\

\begin{figure}
  \centering
  \includegraphics[width=0.45\textwidth]{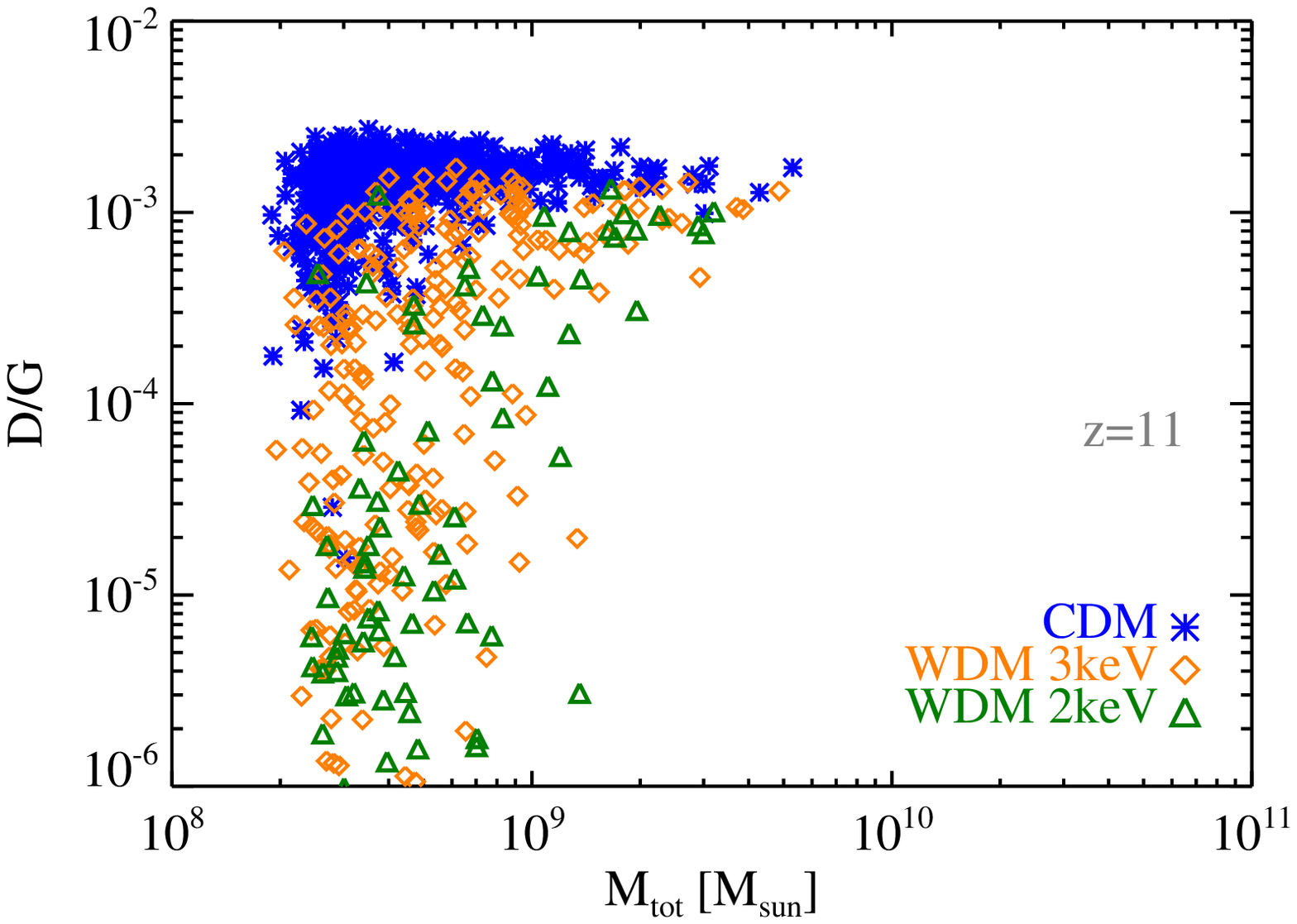} \\
  \includegraphics[width=0.45\textwidth]{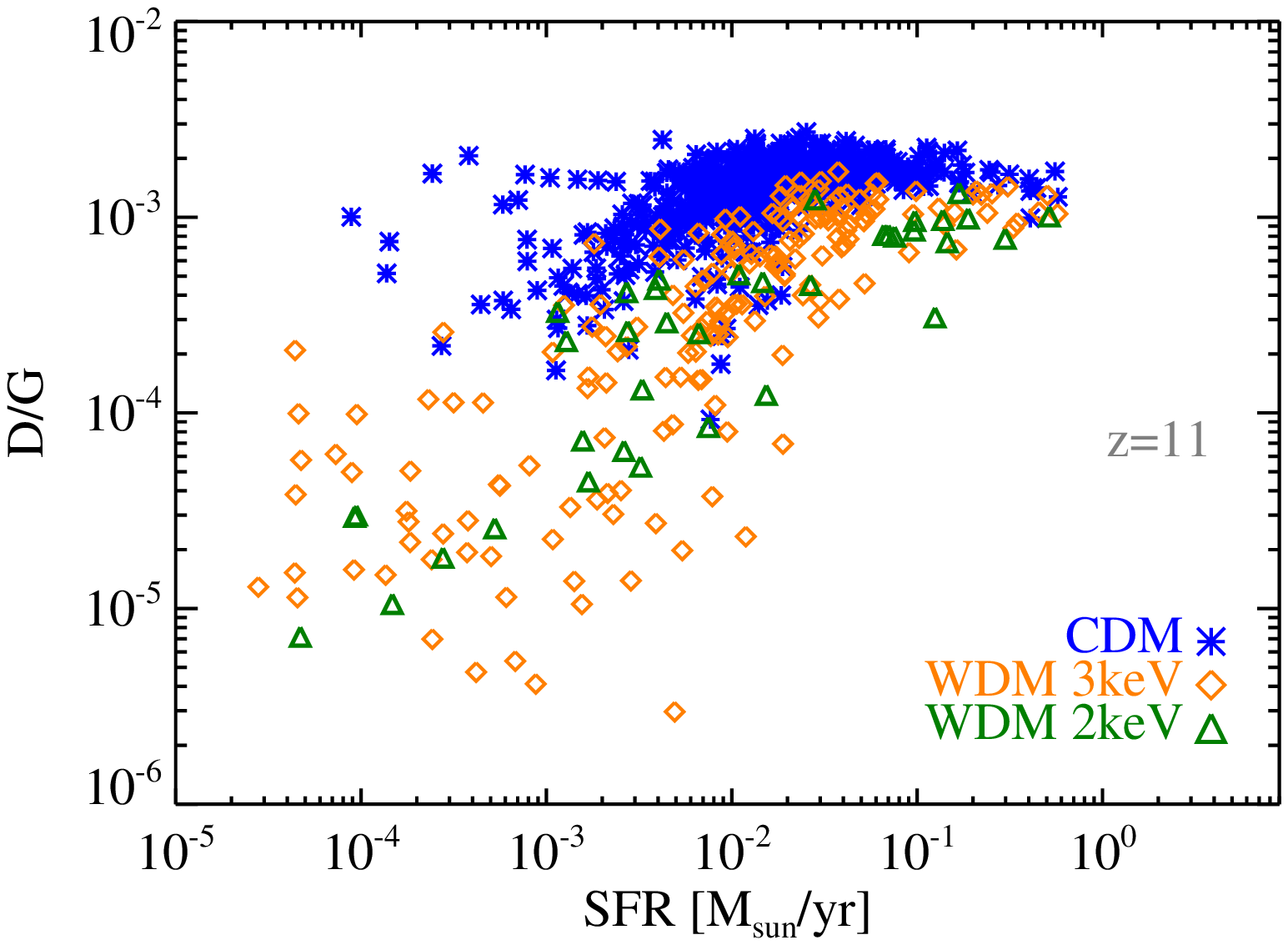}
\caption{Expected dust-to-gas ratios, D/G, as a function of the total halo mass for galaxies in the CDM, 3~keV WDM, and 2~keV WDM fiducial runs (top) and corresponding relation with the local SFR (bottom).
}
  \label{fig:DtoG}
\end{figure}

\noindent
Consistently with the generally low metallicities in early environments, the expected dust content is usually small, as shown for example by the dust-to-gas mass ratios (D/G) in the three fiducial runs at $z\simeq 11$ (see fig.~\ref{fig:DtoG}, where, for the sake of clarity, we have omitted the other simulations).
Despite figures increasing for increasing SFRs (bottom panel), D/G values are much lower than the ones (of the order of $\sim 10^{-2} $) estimated for the local Universe.
The sparsity of data points for small SFRs is consistent with slower and less efficient gas chemical enrichment in primordial WDM haloes, especially for the 2~keV case which experiences the most pronounced effects. 
The implications of the nature of dark matter can be further seen in the paucity of data points and the lower D/G levels reached for a given halo total mass, $ M_{\rm tot} $, in WDM scenarios (upper panel), due to the suppression of small-scale objects and delayed structure growth.
In comparison to the CDM scenario, where D/G values are around $ 10^{-4} - 10^{-3} $ at all masses, the 2~keV WDM model features values down to $ 10^{-6} - 10^{-4} $ at $ M_{\rm tot} < 10^9 \,\rm M_\odot $. 
The effects are milder for 3~keV WDM, in which case the D/G value ranges between $ 10^{-6} $ and $\sim 10^{-3} $.
Masses above $ \sim 10^9 \,\rm M_\odot $ feature converging results for CDM and WDM.\\

\begin{figure*}
  \begin{center}
  \includegraphics[width=0.45\textwidth]{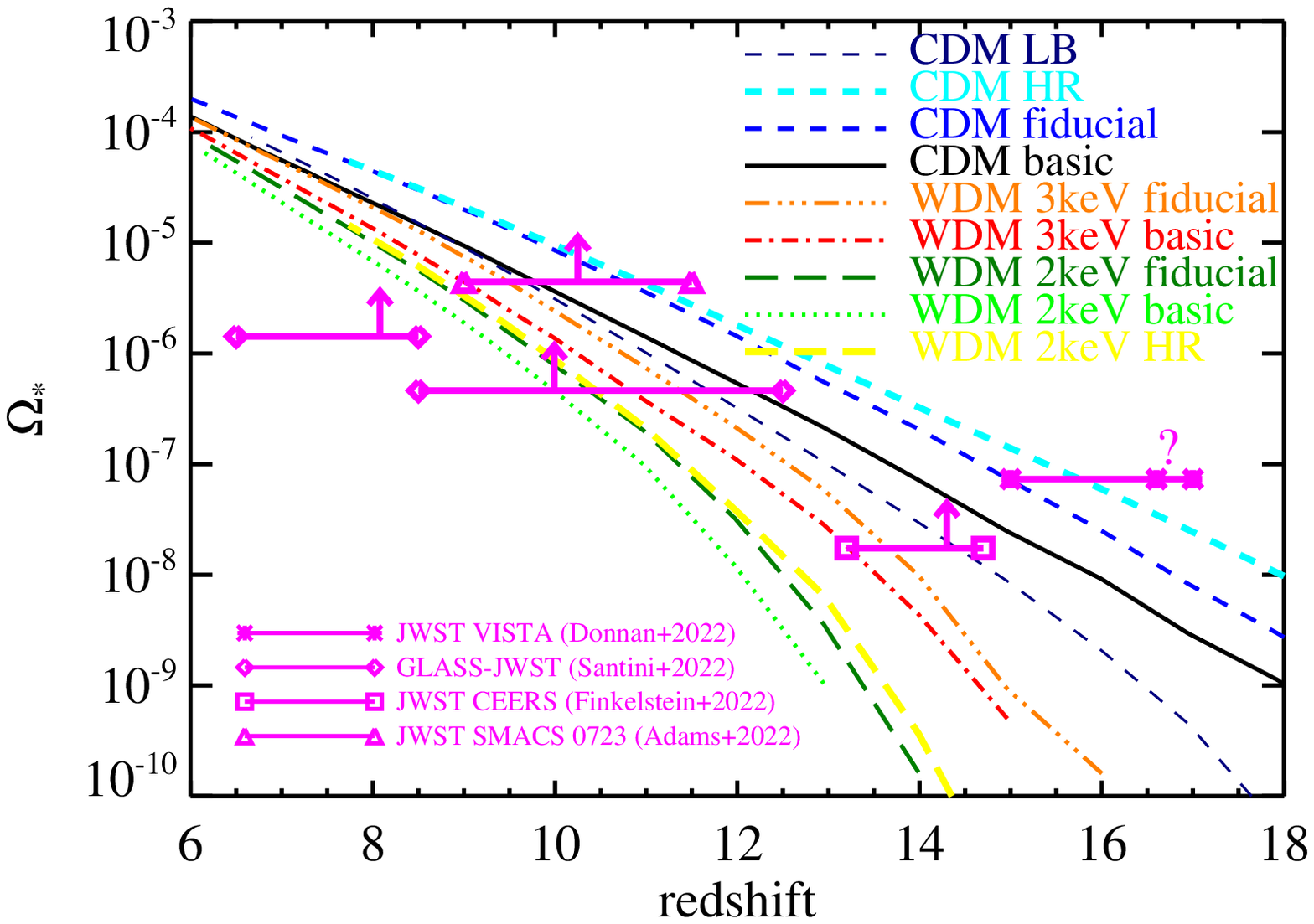} 
  \includegraphics[width=0.45\textwidth]{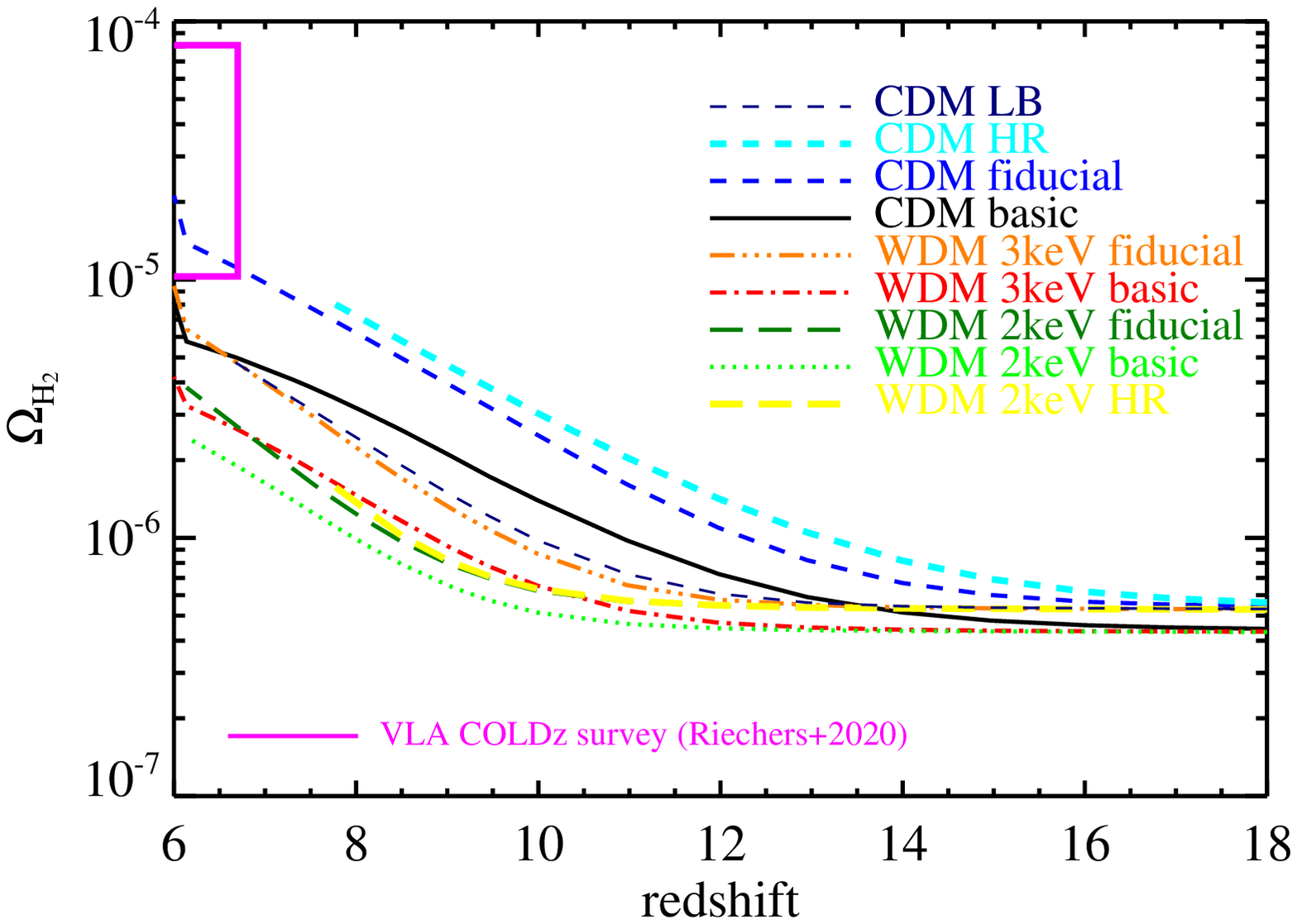} 
\caption{Redshift evolution of the stellar-mass density parameter $\Omega_{\rm \star} $ (left) for different models compared to JWST-inferred values and redshift evolution of the H$_2$ mass density parameter $\Omega_{\rm H_2} $ (right) for different models compared to VLA COLDz data  (magenta box). }
  \label{fig:OmegaStars}
  \end{center}
\end{figure*}

\noindent
Additional physical quantities related to high-$z$ baryon evolution might be precious to shed light on the nature of dark matter.
In fig.~\ref{fig:OmegaStars} the redshift evolution of the cosmological stellar-mass density ($\rho_{\rm \star}$) parameter 
$\Omega_{\rm \star} = \rho_{\rm \star} /\rho_{\rm 0,  crit}$ 
has been plotted for the different numerical simulations considered here and compared to corresponding data at $ 6 \lesssim z \lesssim 16$, as derived from total stellar masses and survey volumes of a number of observational JWST programmes \citep{Santini2022,Adams2022,Finkelstein2022,Donnan2022}.
Since the observed samples are affected by completeness issues, cosmological stellar-mass values shall be considered as lower limits for the actual ones.
By comparing JWST-inferred $\Omega_{\star}$ data and simulation data, we see a generally increasing trend with time, as a consequence of cosmological structure growth (and consistently with the SFRD shown in fig.~\ref{fig:SFRD}).
There are obvious differences in the high-redshift ($ z \gtrsim 10$) window.
While both fiducial and basic CDM simulation results are consistent with JWST data, 2~keV WDM values are below all the $ z \gtrsim 9.5 $ lower limits, independently of the details of the physical implementation or resolution. This is in tension with WDM with particle masses $m_{\rm WDM} \lesssim $~2~keV. For the 3~keV WDM case it is not possible to give definitive assessments, with any conclusion being degenerate with the uncertainties on stellar masses at high $z$ \citep{Santini2022,Finkelstein2022}.
\\
A complementary point of view is given by the redshift evolution of the H$_2$ mass density ($ \rho_{\rm H_2}$) parameter 
$\Omega_{\rm H_2} = \rho_{\rm H_2} /\rho_{\rm 0,crit}$ 
for the different dark-matter models and compared to available constraints at $ z \simeq 6$-7 \citep{Riechers2020}, in the right panel of the figure.
We note that H$_2$ is a powerful tracer of primordial structure formation and a major indicator of cold-gas collapse at all cosmic epochs.
Since it is very sensitive to several physical and chemical processes (gas thermal state, chemical composition, dust content), its investigation is useful to understand the origin of primordial galaxies and rule out non-performing models.
As it is clear from the trends, the fiducial CDM and 3~keV WDM results are in line with VLA constraints, while the 2~keV WDM scenario under-predicts $ \Omega_{\rm H_2} $.
For all the models considered $ \Omega_{\rm H_2} $ evolution is very similar at $ z \gtrsim 14 $, as cosmic gas has not formed significant amounts of molecules yet, and star formation can take place only in a few (larger at the time) haloes.
While the growth of cosmic structures proceed, more and more gas condenses and forms H$_2$, as is visible from the $ \Omega_{\rm H_2} $ behaviour at lower $z$.
\\
It is important to stress that the conclusions about CDM and WDM are based on the detailed implementation employed here. A simpler (basic) modelling would suggest misleading results, with smaller $ \Omega_{\rm H_2} $ values for all dark-matter scenarios and for any $z \gtrsim 7$.
The plot also highlights that changes due to an improved description can be comparable to or larger than the ones induced by a finer resolution, as results, for example, from the trends in the basic, fiducial, and HR CDM/WDM runs.\\

\begin{figure*}
  \centering
  \includegraphics[width=0.45\textwidth]{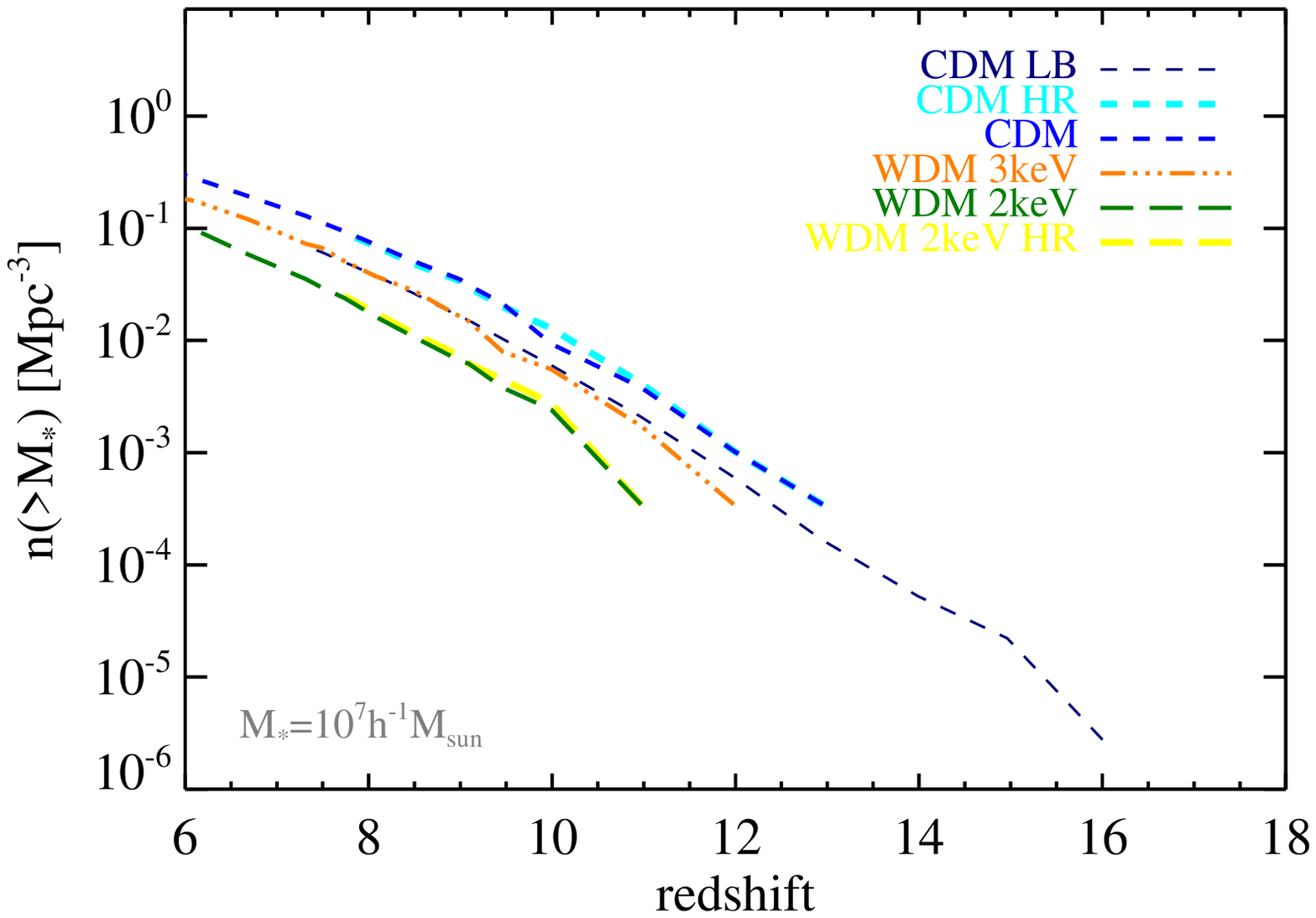}
  \includegraphics[width=0.45\textwidth]{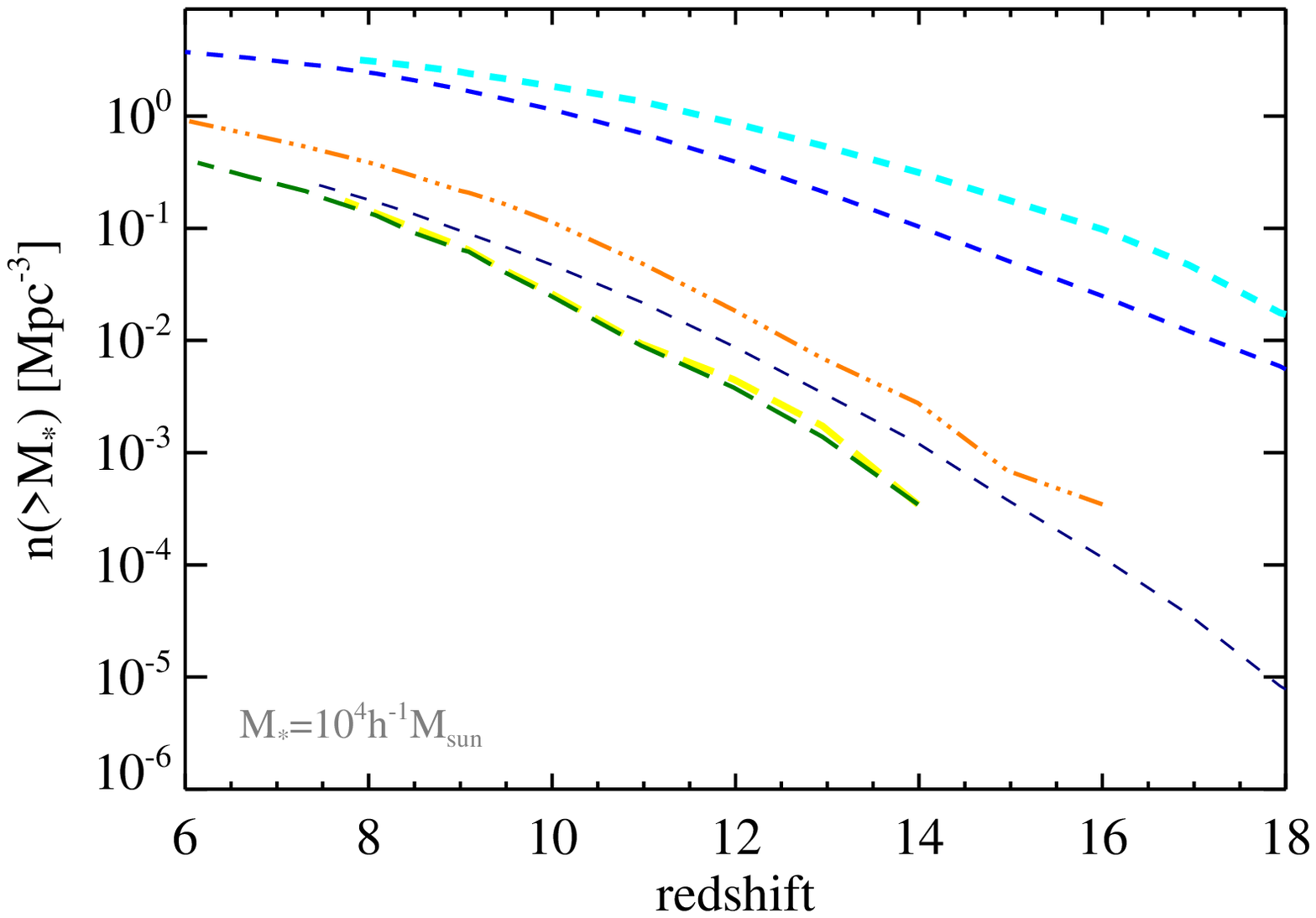}
\caption{Redshift evolution of the expected cumulative number density of objects with a minimum stellar mass of
$ 10^7 \,\rm M_{\odot} / {\it h}$ (left) and $ 10^4 \,\rm M_{\odot} / {\it h}$ (right), 
for CDM, 3~keV WDM, and 2~keV WDM fiducial runs.}
  \label{fig:Nth}
\end{figure*}

\begin{figure*}
  \centering
  \includegraphics[width=0.32\textwidth]{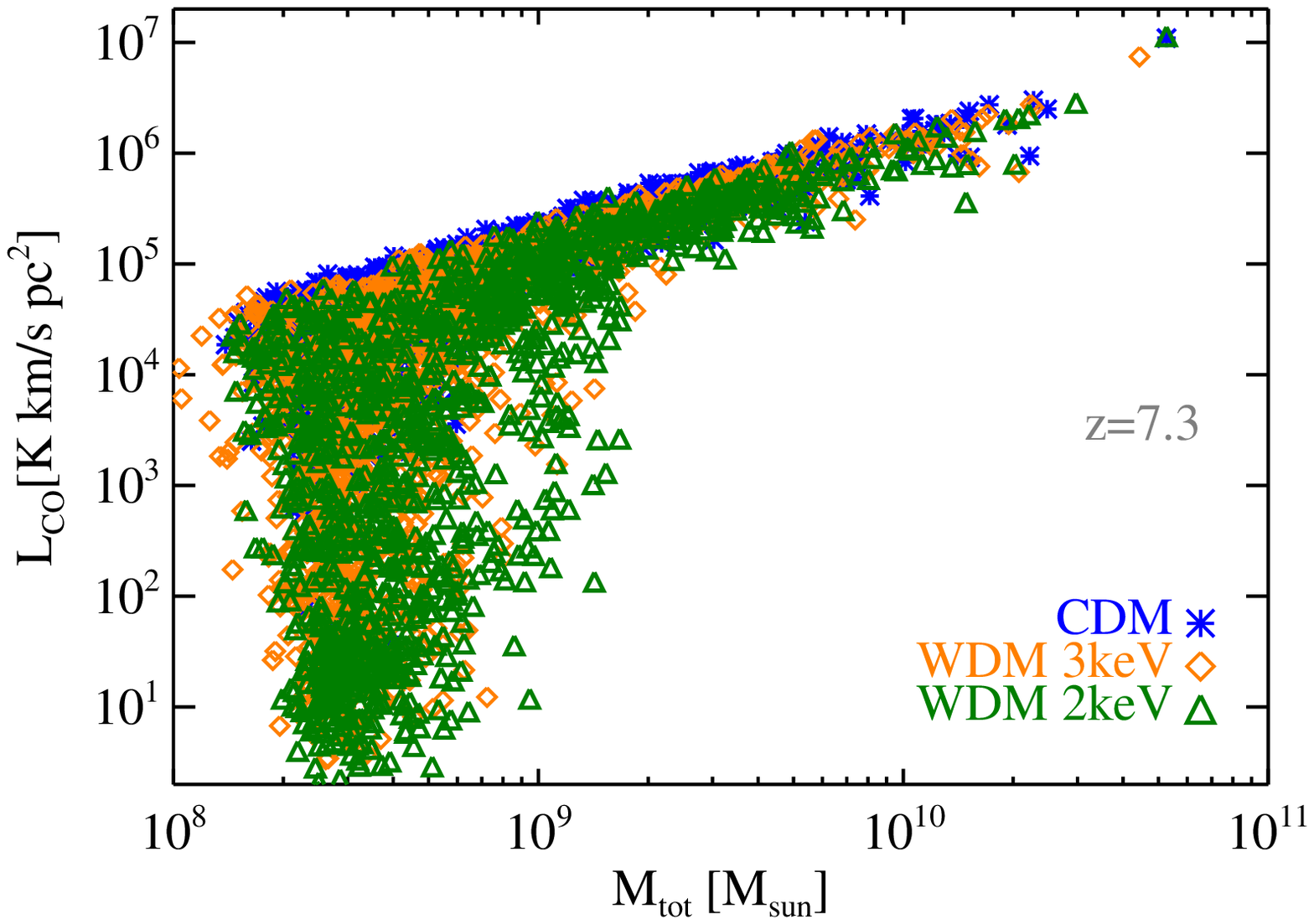}
  \includegraphics[width=0.32\textwidth]{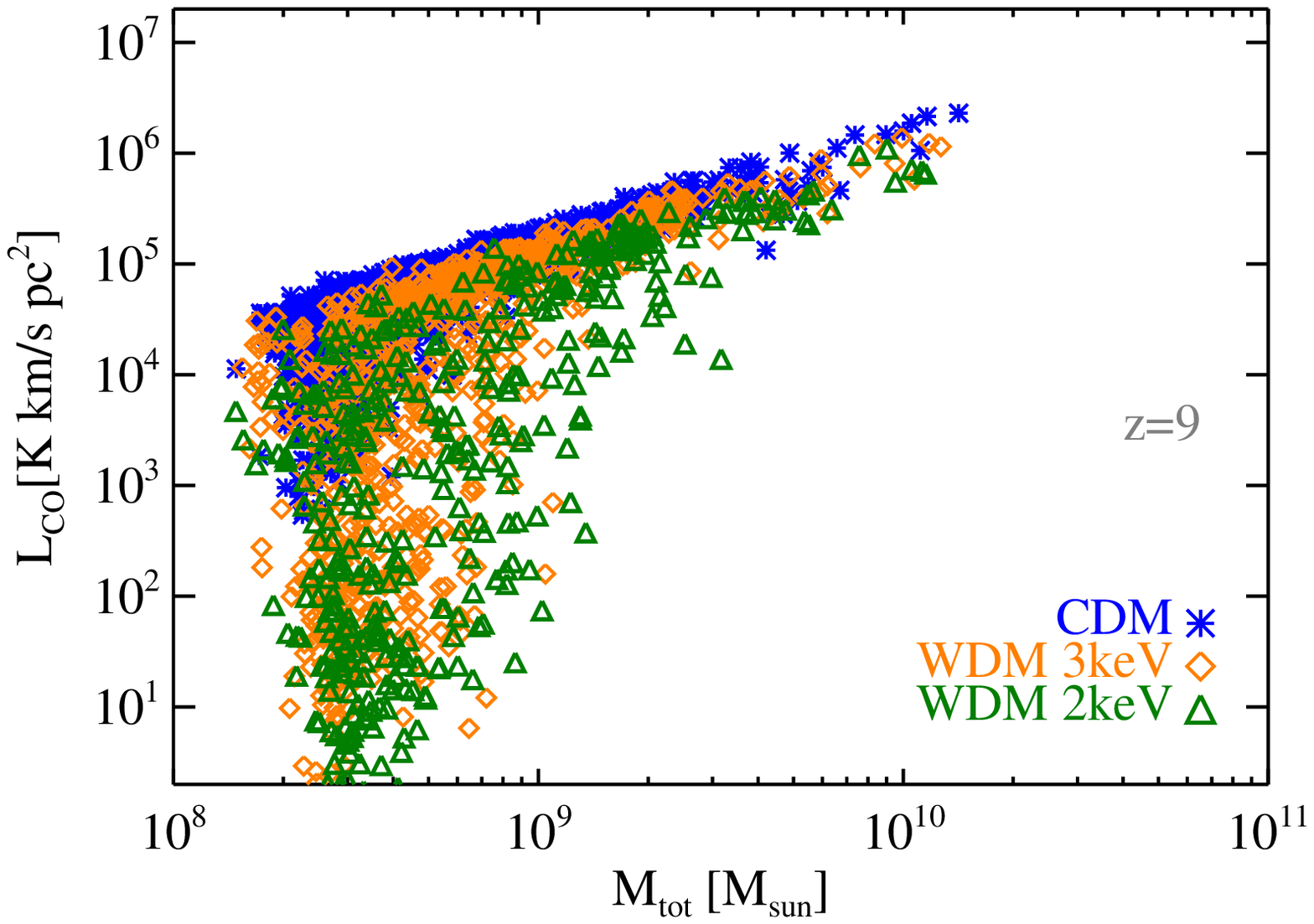}
  \includegraphics[width=0.32\textwidth]{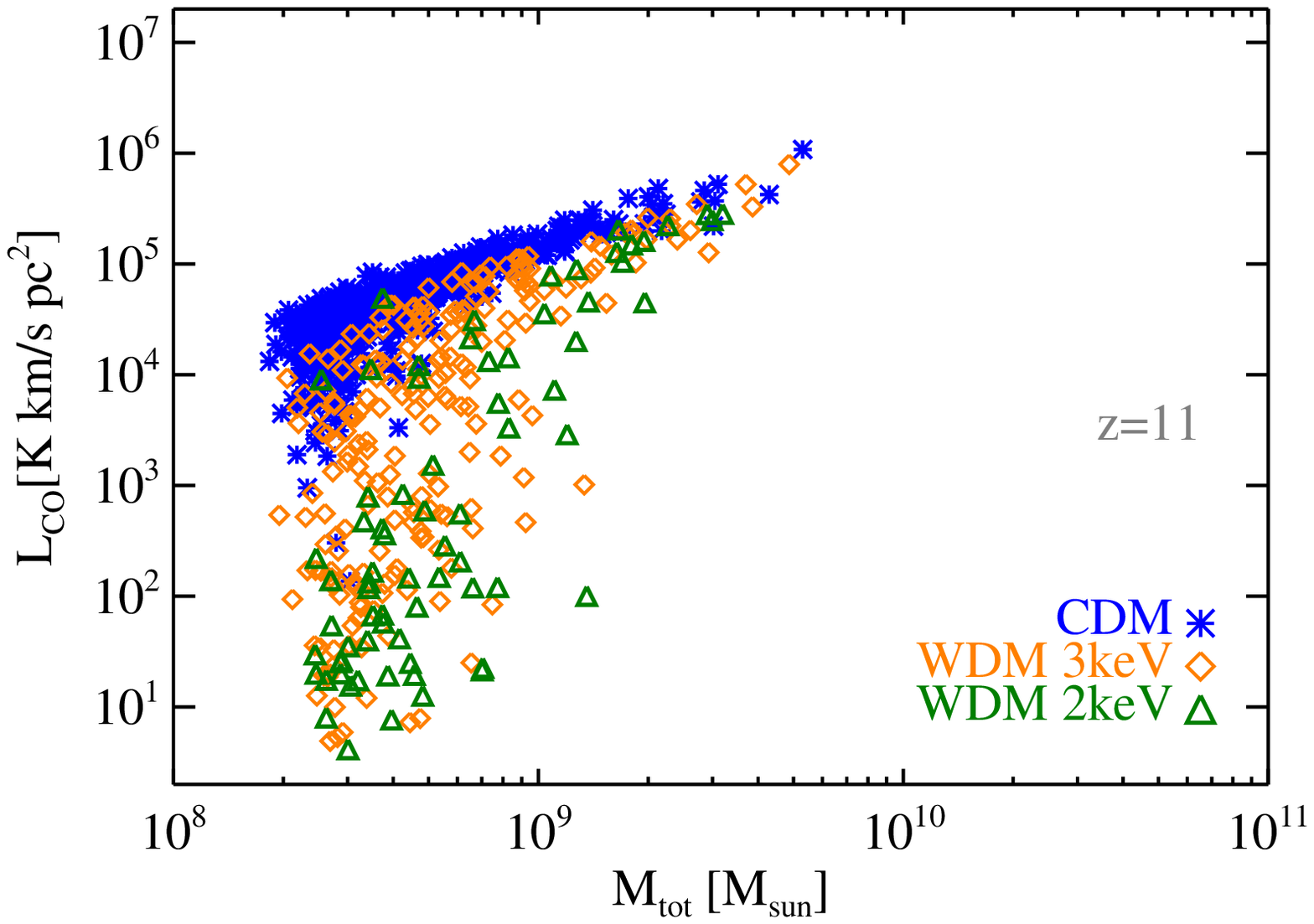}\\
\caption{ Expected $L_{\rm CO} $ emission as function of the total halo mass in the CDM (asterisks), 3~keV WDM (squares), and 2~keV WDM (triangles) fiducial runs at $z = 7.3$ (left), $z = 9$ (centre), and $z = 11$ (right), respectively. }
  \label{fig:Lco}
\end{figure*}

\noindent
Finally, we briefly discuss a couple of possible probes that, in the future, might help disentangle the nature of dark matter.
Information about CDM and WDM implications for cosmic structures could be given by the overall evolution of the cumulative number counts of stellar (potentially visible) objects with mass above a fixed threshold. In fig.~\ref{fig:Nth} we show predicted simulation results in the fiducial CDM and WDM runs for two different stellar-mass thresholds:
$ 10^7\,\rm {M_\odot} / {\it h}$ 
and
$ 10^4\,\rm {M_\odot} / {\it h}$.
As is visible in the figure, the discrepancies of galaxy counts are of about 1 dex at $ z \gtrsim 10 $ when a minimum stellar-mass threshold of $ 10^7\,\rm {M_\odot} / {\it h}$ is assumed (left panel) and reach 2 dex when a more extreme threshold of $ 10^4\,\rm {M_\odot} / {\it h}$ is considered (right panel).
At lower $ z $, differences are smaller, but they can persist down to $ z \simeq 6$.
In the future, once more data will have been collected by JWST and other upcoming facilities, it will be possible to pose more stringent limits, possibly at fraction of keV levels, on different scenarios by exploiting such trends.
\\
\noindent
Another interesting opportunity to pose constraints on dark matter is linked to the observable signatures of early CO emission. 
We note that CO is a strong emitter of galactic gas and its signal has been detected up to $ z \simeq 7$ (with e.g. ALMA, VLA, NOEMA, etc.).
In fig.~\ref{fig:Lco} we show the expected CO luminosity, $ L_{\rm CO} $, as a function of the halo total mass, derived for CDM, 3~keV WDM, and 2~keV WDM cosmologies according to \cite{Bolatto2013}.
For the sake of clarity, we only considered the galaxy populations expected in the fiducial runs at $z \simeq 7.3$, 9, and 11.
The general behaviour is similar in all the models, with scattered points at lower masses ($ \lesssim 10^9\,\rm M_\odot $) and a roughly linear trend as an upper limit.
The scatter is due to the buildup of the molecular content in the haloes that are just hosting gas collapse and primordial star formation, while more evolved structures have already converted a significant fraction of atomic gas into molecules. The latter delimits the $ L_{\rm CO} $-mass relation, which holds for both CDM and WDM galaxies.
Although the physical process is qualitatively the same, typical timescales change.
Indeed, spectral suppression in WDM scenarios induces delays in halo formation and cosmic-gas collapse, during which molecules are formed. 
So, the resulting $ L_{\rm CO} $ signal is tightly bound to the underlying nature of dark matter and the impacts of CDM versus WDM are particularly visible at early times.
WDM small-scale suppression is recognizable at $ z \gtrsim 9$, when, below $ 10^9 \,\rm M_\odot $, there is a deficit in the expected $ L_{\rm CO} $ emission up to a few dex (right panel at $ z = 11$).
At that time, CDM galaxies have already had enough time to complete molecule formation, while 2~keV WDM galaxies feature poorer number statistics and lower molecular content. The 3~keV WDM case expects galaxies that are in more advanced stages, but that have not reached the CDM ones, yet.
By the first half billion years ($z = 9$, central panel), the trends start to converge and at $z = 7.3$ (left panel) different models are basically indistinguishable, since the original differences between CDM and WDM are erased by the ongoing feedback effects.
This catchup takes place in only a few hundreds of million years.
For this reason although challenging, tight constraints on the nature of dark matter might come through detection of CO emission at very early times.\\
%

%*****************************************************************************

\section{Discussion and conclusions} \label{sect:conclusions}

\noindent
In this work we have exploited the latest JWST observational determinations and novel up-to-date numerical simulations to put constraints on the nature of dark matter from high-redshift observations.
We have compared the latest JWST-inferred high-$z$ star formation estimates 
\citep{Santini2022, Donnan2022, Finkelstein2022, Adams2022,Harikane2022, Bouwens2022} 
with a set of non-equilibrium hydrodynamical simulations which incorporate the new, rich, and accurate modelling of cosmic structure formation at early times by \cite{Maio2022}.
This attempt is the first one to try to set constraints on WDM by combining such modelling with state-of-the-art JWST observations at extremely high redshift.
Previous works based on high-redshift hydro-simulations have either neglected a fully complete modelling of primordial gas and structures in CDM and WDM or had no or little data support for the primordial regimes probed by JWST.\\
We contrast galaxy buildup in the standard CDM model against two models with 2 and 3~keV WDM, respectively.
We adopted cosmological matter density and expansion parameters that are consistent with both the standard model and WMAP data.
The latest Planck measurements \citep{Planck2014, Planck2020} suggest slightly different values, while spectral parameters are similar.
This is an important point, because early structure formation and halo mass functions are mostly affected by variations in $\sigma_8$ which is consistently constrained by the aforementioned experiments.
Thus, changing the initial parameter set does not lead to appreciable differences in our results and the overall trends are preserved \cite[see also discussions in e.g.][]{Maio2010, Maio2011}.\\
We generated initial conditions at high redshift via cosmological linear perturbation theory, which is well suited for such early regimes.
Coherent supersonic flows of the baryons relative to the underlying dark-matter distribution on megaparsec scales are caused by higher-order corrections accounting for the advection of small-scale perturbations by large-scale velocity flows after decoupling.
However, we have verified that, independently from the initial redshift (ranging between $z=100$ and $z=1020$), the implications of such bulk motions have no impact on the masses in the epoch studied in this work \citep{Maio2011bulk}.\\
As is typically done, we assumed that the statistical distribution of the primordial matter perturbation field is Gaussian.
Deviations from Gaussianity could be present and could enhance or dampen the occurrence of objects with a given mass; nevertheless, the expected level of these primordial non-Gaussianities is so small that possible implications on molecular evolution, popIII and popII-I  star formation, metal enrichment, gas temperature, and possibly detectable signals would be negligible and dominated by baryon effects
\citep{MaioIannuzzi2011, Maio2011ng, MaioKhochfar2012, Maio2012ngGRB}.
\\
More critical uncertainties are about feedback processes and their degeneracies with the nature of dark matter.
Fortunately, feedback effects usually have local impacts and alter the local chemical and physical properties of cosmic objects.
Although their efficiency is poorly constrained, they play a significant role at low $z$, when structure evolution is in more advanced stages and possible dark-matter signatures have already been washed out \citep{Schneider2014}.
Primordial galaxies are young structures and have experienced little feedback effects; therefore, their statistical occurrence is mainly driven by the underlying dark-matter model.
For this reason, the early Universe is a precious window to test dark-matter models and, furthermore, calibrating feedback parameters in the low-$z$ regime together with large high-$z$ data samples might make it possible to both break the degeneracies and to provide hints on the late-time evolution of cosmic structures.\\
Predicted stellar and molecular mass density parameters, as well as star formation rate densities are consistent with early results of JWST data at $ z \gtrsim 7$ and with previous VLA constraints at $ z \simeq 7 $ in the CDM and 3~keV WDM scenarios.
JWST data do not show any hint of an excess of number densities of bright galaxies at $ z \gtrsim 7 $, compared to the standard model.
Thus, current data are neither in tension with cold dark matter nor warm dark matter models with $m_{\rm WDM} > $~2~keV.
Current data are mainly probing young, bright and rare objects, whose physical properties are remarkably similar in the different scenarios. 
Consistently with the short cosmic time of the infant Universe, these are expected to host little dust content.
However, due to the low-mass power suppression, the faint end of the UV  luminosity function flattens for lighter dark-matter particles, while the corresponding UV correlation function steepens significantly.
These two different observables, especially when used in combination, can be extremely promising not only for constraining galaxy formation models \citep{vandaalen16}, but also for disentangling dark-matter scenarios by using faint high-$z$ visible sources.
In the future, when more data for small, dim, high-$z$ sources will be available, further hints may come from number density statistics and galactic CO emission luminosities.

%*****************************************************************************

% =============================================================================

\begin{acknowledgements}
We warmly thank the anonymous referee for his/her prompt feedback which helped us improve the manuscript and extend the original text with extremely constructive comments.
UM acknowledges useful discussion with M.~Castellano and is thankful to P.~Santini and C.~Donnan for sharing preliminary analysis of early JWST data. MV is supported by the PD51-INFN INDARK and ASI-INAF n. 2017-14-H.0 grants.
\end{acknowledgements}

% =============================================================================

\bibliographystyle{aa1}
\bibliography{bibl}

% =============================================================================

%\appendix
% \section{Title}  \label{AppendixData}

% =============================================================================

\end{document}